\documentclass[12pt,a4paper]{article}
	
\usepackage{a4wide}
\usepackage{amsmath}
\usepackage{amssymb}
\usepackage{amsfonts}
\usepackage{epsfig}
\usepackage{exscale}
\usepackage{float}
\usepackage{bbm}
\usepackage[numbers,sort&compress]{natbib}

\newcommand{\R}{{\mathbb{R}}}
\newcommand{\C}{{\mathbb{C}}}

\newcommand{\1}{1\!\!1}
\newcommand{\0}{0\hspace{-1.65mm}0}
\newcommand{\II}{\rm I\hspace{-0.3mm}I}
\newcommand{\I}{\rm I}
\newcommand{\p}{\partial}

\makeatletter
\@addtoreset{equation}{section}
\makeatother

\title{From a Particle in a Box to the \\ 
Uncertainty Relation in a Quantum Dot \\
and to Reflecting Walls for Relativistic Fermions}

\author{M.\ H.\ Al-Hashimi and U.-J.\ Wiese \\ \\
Albert Einstein Center for Fundamental Physics \\
Institute for Theoretical Physics, Bern University \\
Sidlerstrasse 5, CH-3012 Bern, Switzerland \\ \\}

\begin{document} 

\maketitle

\vspace{-1cm}

\begin{abstract} \normalsize

We consider a 1-parameter family of self-adjoint extensions of the Hamiltonian 
for a particle confined to a finite interval with perfectly reflecting boundary
conditions. In some cases, one obtains negative energy states which seems to
violate the Heisenberg uncertainty relation. We use this as a motivation to
derive a generalized uncertainty relation valid for an arbitrarily shaped 
quantum dot with general perfectly reflecting walls in $d$ dimensions. In 
addition, a general uncertainty relation for non-Hermitean operators is derived 
and applied to the non-Hermitean momentum operator in a quantum dot. We also 
consider minimal uncertainty wave packets in this situation, and we prove that
the spectrum depends monotonically on the self-adjoint extension parameter. In
addition, we construct the most general boundary conditions for semiconductor 
heterostructures such as quantum dots, quantum wires, and quantum wells, which
are characterized by a 4-parameter family of self-adjoint extensions.  
Finally, we consider perfectly reflecting boundary conditions for relativistic 
fermions confined to a finite volume or localized on a domain wall, which are 
characterized by a 1-parameter family of self-adjoint extensions in the 
$(1+1)$-d and $(2+1)$-d cases, and by a 4-parameter family in the $(3+1)$-d and 
$(4+1)$-d cases.

\end{abstract}

\newpage
 
\section{Introduction}

The subtle differences between Hermiticity and self-adjointness of quantum
mechanical operators, which were first understood by von Neumann \cite{Neu32},
are rarely emphasized in quantum mechanics textbooks. This already affects the 
elementary textbook problem of a particle in a box \cite{Bon01}. Almost
exclusively, the students are taught to set the wave function to zero at the
boundary, in order to ensure its continuity. However, it is sufficient to
guarantee that no probability leaks outside the box, i.e.\ that the probability 
current (but not necessarily the wave function itself) vanishes at the boundary.
The resulting most general perfectly reflecting boundary condition contains a
real-valued parameter that characterizes a family of self-adjoint extensions of 
the quantum mechanical Hamiltonian. In general, the wave function then does not 
go to zero at the boundary, and consequently the probability density to find the
particle directly at the wall does not vanish. This is consistent with 
classical intuition of a ball bouncing off a perfectly reflecting wall. 

While the self-adjoint extensions of the Hamiltonian are certainly known to the 
experts, in the beginning of this paper we introduce them in a pedagogical
manner, since they, unfortunately, seem not to constitute common knowledge in
quantum mechanics. When re-doing the standard textbook problem of a particle in
a box, we will find states of negative energy for a particle with only kinetic
energy. Such states seem to violate the Heisenberg uncertainty relation. In 
order to clarify this issue we then derive a generalized uncertainty relation 
valid for a particle confined to an arbitrarily shaped region with general 
perfectly reflecting walls in $d$ dimensions. This situation is relevant in the 
context of quantum dots. The most general perfectly reflecting boundary 
condition ensures that the component of the current normal to the reflecting 
surface must vanish. Again there is a 1-parameter family of self-adjoint 
extensions. The real-valued parameter that characterizes the boundary condition 
is a material-specific constant whose value could be determined experimentally 
for actual quantum dots. In particular, this parameter enters the generalized 
uncertainty relation. We also consider minimal uncertainty wave packets in a 
general quantum dot. Interestingly, when the self-adjoint extension parameter 
vanishes, a constant wave function of zero energy has $\Delta p = 0$ and 
saturates the generalized uncertainty relation. While the purpose of our paper 
is to some extent pedagogical, to the best of our knowledge the generalized 
uncertainty relation for quantum dots has not been derived before. Perfectly
reflecting boundary conditions for relativistic fermions have been investigated
in the context of the MIT bag model \cite{Cho74,Cho74a,Has78}. Here we construct
the most general boundary condition for relativistic Dirac fermions, which is 
characterized by a 1-parameter family of self-adjoint extensions in the 
$(1+1)$-d case, and by a 4-parameter family in the $(3+1)$-d case. Finally, we 
extend the discussion to domain wall fermions in $(2+1)$-d and $(4+1)$-d 
space-times.

The paper is organized as follows. In section 2 we discuss a particle confined
to a finite interval with general perfectly reflecting boundary conditions, and
we derive the corresponding generalized uncertainty relation. In section 3 this
relation is extended to an arbitrarily shaped quantum dot with perfectly
reflecting walls in $d$ dimensions. We also derive a general uncertainty 
relation for non-Hermitean operators and apply it to the non-Hermitean momentum
operator. In addition, we construct the corresponding most general minimal 
uncertainty wave packet, and we prove that the spectrum varies monotonically
with the self-adjoint extension parameter. We also construct the most general 
boundary conditions for semiconductor heterostructures. In section 4, we extend 
the discussion to relativistic Dirac fermions in $(1+1)$-d and $(3+1)$-d, and in
section 5 to domain wall fermions in $(2+1)$-d and $(4+1)$-d space-times.
Finally, section 6 contains our conclusions.

\section{Particle in a 1-d Box with General Perfectly Reflecting Walls}

Let us consider a particle of mass $m$ moving in the 1-d interval 
$\Omega = [-L/2,L/2]$. This problem has been discussed in the context of 
self-adjoint extensions in \cite{Bon01}. Other examples of quantum mechanical
problems involving the theory of self-adjoint extensions are discussed, for
example, in \cite{Ara04}. We use natural units in which $\hbar = 1$. For 
simplicity, we restrict the Hamiltonian to the kinetic energy operator
\begin{equation}
H = \frac{p^2}{2 m} = - \frac{1}{2 m} \p_x^2.
\end{equation}
The wave function $\Psi(x,t)$ gives rise to the probability current density 
\begin{equation}
\label{current}
j(x,t) = \frac{1}{2 m i} [\Psi(x,t)^* \p_x \Psi(x,t) - 
\p_x \Psi(x,t)^* \Psi(x,t)], 
\end{equation}
which together with the probability density $\rho(x,t) = |\Psi(x,t)|^2$ obeys 
the continuity equation
\begin{equation}
\p_t \rho(x,t) + \p_x j(x,t) = 0.
\end{equation}
In the following discussion, we can ignore the time-dependence of the wave 
function and simplify the notation to $\Psi(x)$.

\subsection{Spatial Boundary Conditions}

In order to guarantee probability conservation, one must demand that the 
probability current vanishes at the boundary, i.e.\
\begin{equation}
j(L/2) = j(-L/2) = 0.
\end{equation}
The most general local boundary condition that implies this takes the form
\begin{equation}
\gamma(x) \Psi(x) + \p_x \Psi(x) = 0, \quad x = \pm L/2.
\end{equation}
Indeed, one then obtains
\begin{equation}
j(x) = \frac{1}{2 m i} [- \Psi(x)^* \gamma(x) \Psi(x) + 
\gamma(x)^* \Psi(x)^* \Psi(x)] = 0 \ \Rightarrow \ \gamma(x) \in \R, \quad
x = \pm L/2.
\end{equation}
The two real-valued parameters $\gamma(L/2)$ and $\gamma(-L/2)$ characterize a 
1-parameter family of self-adjoint extensions of the Hamiltonian at each of the 
two ends of the interval $\Omega$. In order not to break parity via the boundary
conditions, we restrict ourselves to
\begin{equation}
\gamma(L/2) = - \gamma(-L/2) = \gamma \in \R,
\end{equation}
such that
\begin{equation}
\label{bc}
\gamma \Psi(L/2) + \p_x \Psi(L/2) = 0, \quad
- \gamma \Psi(-L/2) + \p_x \Psi(-L/2) = 0.
\end{equation}

\subsection{Self-Adjointness of the Hamiltonian}

In order to investigate whether the Hamiltonian is indeed self-adjoint when the 
wave functions obey the boundary conditions eq.(\ref{bc}), we now consider
\begin{eqnarray}
\langle\chi|H|\Psi\rangle&=&
- \frac{1}{2 m} \int_{-L/2}^{L/2} dx \ \chi(x)^* \p_x^2 \Psi(x) \nonumber \\
&=&\frac{1}{2 m} \int_{-L/2}^{L/2} dx \ \p_x \chi(x)^* \p_x \Psi(x) - 
\frac{1}{2 m} \left[\chi(x)^* \p_x \Psi(x)\right]_{-L/2}^{L/2} \nonumber \\
&=&- \frac{1}{2 m} \int_{-L/2}^{L/2} dx \ \p_x^2 \chi(x)^* \Psi(x) + 
\frac{1}{2 m} 
\left[\p_x \chi(x)^* \Psi(x) - \chi(x)^* \p_x \Psi(x)\right]_{-L/2}^{L/2} 
\nonumber \\
&=&\langle\Psi|H|\chi\rangle^* + \frac{1}{2 m}
\left[\p_x \chi(x)^* \Psi(x) - \chi(x)^* \p_x \Psi(x)\right]_{-L/2}^{L/2}.
\end{eqnarray}
The Hamiltonian is Hermitean (or symmetric in mathematical parlance) if
\begin{equation}
\langle\chi|H|\Psi\rangle = \langle H^\dagger \chi|\Psi\rangle =
\langle H \chi|\Psi\rangle = \langle\Psi|H|\chi\rangle^*,
\end{equation}
which is indeed the case if
\begin{equation}
\label{symmetricH}
\left[\p_x \chi(x)^* \Psi(x) - \chi(x)^* \p_x \Psi(x)\right]_{-L/2}^{L/2} = 0.
\end{equation}
The domain $D(H)$ of the Hamiltonian contains the at least twice-differentiable 
square-integrable wave functions $\Psi(x)$ that obey the boundary condition 
eq.(\ref{bc}). Using that condition, eq.(\ref{symmetricH}) reduces to
\begin{equation}
\Psi(L/2) \left[\p_x \chi(L/2)^* + \gamma \chi(L/2)^*\right] -
\Psi(-L/2) \left[\p_x \chi(- L/2)^* - \gamma \chi(-L/2)^*\right] = 0.
\end{equation}
Since $\Psi(L/2)$ and $\Psi(-L/2)$ can take arbitrary values, the Hamiltonian
is Hermitean if
\begin{equation}
\gamma \chi(L/2) + \p_x \chi(L/2) = 0, \quad
- \gamma \chi(-L/2) + \p_x \chi(-L/2) = 0,
\end{equation}
i.e.\ if the wave function $\chi(x)$ also obeys the boundary condition
eq.(\ref{bc}). Imposing this boundary condition also on $\chi(x)$ implies that
the domain of $H^\dagger$ coincides with the domain of $H$, 
$D(H^\dagger) = D(H)$. Since $H$ is indeed Hermitean when both $\Psi(x)$ and
$\chi(x)$ obey eq.(\ref{bc}), and since, in addition, $D(H^\dagger) = D(H)$, the
Hamiltonian is, in fact, self-adjoint. 

It should be noted that there is even a 4-parameter family of self-adjoint 
extensions of $H$ \cite{Car90}. Here we have encountered only two parameters, 
$\gamma(L/2)$ and $\gamma(-L/2)$, which we have reduced to one parameter 
$\gamma$ by demanding parity symmetry. The other two parameters of the 
4-parameter family of self-adjoint extensions relate the values of the wave 
function and its derivative at $x = L/2$ to the corresponding values at 
$x = - L/2$. Such a boundary condition violates locality and is thus not 
physically meaningful in the present context.

\subsection{Non-Hermiticity of the Momentum Operator}

In order to investigate whether the momentum operator $p = - i \p_x$ is
self-adjoint or at least Hermitean, let us also consider
\begin{eqnarray}
\langle\chi|p|\Psi\rangle&=&- i \int_{-L/2}^{L/2} dx \ \chi(x)^* \p_x \Psi(x)
\nonumber \\
&=&i \int_{-L/2}^{L/2} dx \ \p_x \chi(x)^* \Psi(x) - 
i \left[\chi(x)^* \Psi(x)\right]_{-L/2}^{L/2} \nonumber \\
&=&\langle\Psi|p|\chi\rangle^* - i \left[\chi(x)^* \Psi(x)\right]_{-L/2}^{L/2}.
\end{eqnarray}
Hence, $p$ would be Hermitean only if
\begin{equation}
\label{symmetricp}
\chi(L/2)^* \Psi(L/2) = \chi(-L/2)^* \Psi(-L/2).
\end{equation}
There is no reason why this should be the case when $\Psi(x)$ and $\chi(x)$
obey the boundary condition eq.(\ref{bc}). Hence, in the domain $D(H)$ of the
Hamiltonian, the momentum operator $p$ is not even Hermitean, and thus certainly
not self-adjoint. 

The only self-adjoint extension of the momentum operator on a finite interval 
results from the boundary condition
\begin{equation}
\label{bcpsi}
\Psi(L/2) = \lambda \Psi(-L/2), \quad \lambda \in \C.
\end{equation}
Inserting this relation in eq.(\ref{symmetricp}), we obtain
\begin{equation}
\label{bcphi}
\chi(L/2)^* \lambda \Psi(-L/2) = \chi(-L/2)^* \Psi(-L/2) \ \Rightarrow \ 
\chi(L/2) = \frac{1}{\lambda^*} \chi(-L/2).
\end{equation}
If $\Psi(x)$ obeys eq.(\ref{bcpsi}) and $\chi(x)$ obeys eq.(\ref{bcphi}) the
operator $p$ is Hermitean (i.e.\ symmetric). The domain $D(p)$ contains those
at least once-differentiable square-integrable wave functions $\Psi(x)$ that
obey eq.(\ref{bcpsi}). The domain $D(p^\dagger)$, on the other hand, contains
the corresponding wave functions $\chi(x)$ that obey eq.(\ref{bcphi}). The
operator $p$ is self-adjoint only if $D(p) = D(p^\dagger)$ which implies
\begin{eqnarray}
\label{bcp}
&&\lambda = \frac{1}{\lambda^*} \ \Rightarrow \ \lambda = \exp(i \theta) \ 
\Rightarrow 
\nonumber \\
&&\Psi(L/2) = \exp(i \theta) \Psi(-L/2), \quad
\chi(L/2) = \exp(i \theta) \chi(-L/2).
\end{eqnarray}
Hence, the momentum operator is self-adjoint only if the probability density
is a periodic function, i.e.\ $\rho(L/2) = |\Psi(L/2)|^2 = |\Psi(-L/2)|^2 =
\rho(-L/2)$. Since non-local periodic boundary conditions make no physical 
sense in the present context, and since the wave functions in the domain of the 
self-adjoint Hamiltonian obey the boundary condition eq.(\ref{bc}), but not 
eq.(\ref{bcp}), in this case the momentum operator is neither Hermitean nor 
self-adjoint. Hence, we must conclude that, in the present context, momentum is 
not a physical observable. This indeed makes sense for a particle that is 
confined to a finite region of space. After all, a momentum measurement would 
put the particle in a momentum eigenstate, which also exists outside the box 
and would therefore require infinite energy. An alternative point of view is 
taken in \cite{Gar04,Dia07} where the infinite potential in the energetically 
forbidden region is approached as a limit of a large but finite potential.

\subsection{Energy Spectrum and Energy Eigenstates}

Let us now consider the energy eigenstates and the corresponding energy
eigenvalues for the particle in a box with general reflecting boundary
conditions parameterized by $\gamma$. First, we consider positive energy states
of even parity, i.e.\
\begin{equation}
\Psi_n(x) = A \cos(k_n x), \ E_n = \frac{k_n^2}{2 m}, \ n = 0,2,4,\dots,\infty.
\end{equation}
The boundary condition eq.(\ref{bc}) then implies
\begin{equation}
\gamma \cos(k_n L/2) - k_n \sin(k_n L/2) = 0 \ \Rightarrow \
\frac{\gamma}{k_n} = \tan(k_n L/2).
\end{equation}
Interestingly, for $\gamma = 0$ there is a zero-energy solution with $k_n = 0$
and a constant wave function $\Psi_0(x) = \sqrt{1/L}$. 

Similarly, the positive energy states of odd parity take the form
\begin{equation}
\Psi_n(x) = A \sin(k_n x), \ E_n = \frac{k_n^2}{2 m}, \ n = 1,3,5,\dots,\infty,
\end{equation}
and must obey
\begin{equation}
\gamma \sin(k_n L/2) + k_n \cos(k_n L/2) = 0 \ \Rightarrow \
\frac{\gamma}{k_n} = - \cot(k_n L/2).
\end{equation}
In this case, a zero-energy solution exists for $\gamma = - 2/L$ with the wave
function $\Psi_1(x) = \sqrt{12/L^3} \ x$. While this solution emerges from
$\Psi_1(x) = A \sin(k_1 x)$ in the limit $\gamma \rightarrow - 2/L$, it also
follows directly from the zero-energy Schr\"odinger equation 
$\p_x^2 \Psi(x) = 0$, and the boundary condition 
$(- 2/L) \Psi(L/2) + \p_x \Psi(L/2) = 0$.

Next, let us consider eigenstates of negative energy and even parity. In that
case, the wave function takes the form
\begin{equation}
\Psi_0(x) = A \cosh(\varkappa x), \ E_0 = - \frac{\varkappa^2}{2 m},
\end{equation}
and must obey
\begin{equation}
\gamma \cosh(\varkappa L/2) + \varkappa \sinh(\varkappa L/2) = 0 \ \Rightarrow \
\frac{\gamma}{\varkappa} = - \tanh(\varkappa L/2).
\end{equation}
Again, for $\gamma = 0$ one recovers the zero-energy state 
$\Psi_0(x) = \sqrt{1/L}$.

Finally, we consider the negative energy eigenstates with odd parity, i.e.
\begin{equation}
\Psi_1(x) = A \sinh(\varkappa x), \ E_1 = - \frac{\varkappa^2}{2 m},
\end{equation}
which must obey
\begin{equation}
\gamma \sinh(\varkappa L/2) + \varkappa \cosh(\varkappa L/2) = 0 \ \Rightarrow \
\frac{\gamma}{\varkappa} = - \coth(\varkappa L/2).
\end{equation}
In this case, for $\gamma = - 2/L$ we recover the zero-energy eigenstate
$\Psi_1(x) = \sqrt{12/L^3} \ x$.

The spectrum as a function of $\gamma$ as well as the corresponding wave 
functions are illustrated in figure 1. For $\gamma = \infty$, we recover the
standard textbook case with $\Psi(L/2) = \Psi(-L/2) = 0$. In that case, the
energy spectrum takes the familiar form
\begin{equation}
E_n(\gamma \rightarrow \infty) = \frac{\pi^2 (n + 1)^2}{2 m L^2}, \quad 
n = 0,1,2,\dots,\infty.
\end{equation}
As $\gamma$ decreases to zero, the energy eigenvalues decrease such that
\begin{equation}
E_n(\gamma = 0) = \frac{\pi^2 n^2}{2 m L^2}, \quad n = 0,1,2,\dots,\infty.
\end{equation}
For $- 2/L < \gamma < 0$ there is one negative energy state, and for 
$\gamma < - 2/L$ there are even two negative energy states, which reach negative
infinite energies in the limit $\gamma \rightarrow - \infty$, i.e.
\begin{equation}
E_{0,1}(\gamma \rightarrow - \infty) \rightarrow - \frac{\gamma^2}{2 m} 
\rightarrow - \infty.
\end{equation} 
In the limit $\gamma \rightarrow - \infty$, the wave functions of the negative
energy states reduce to $\delta$-function-like structures localized at the 
boundaries. The rest of the spectrum is exactly as in the standard textbook 
case (i.e.\ $\gamma = \infty$), namely
\begin{equation}
E_n(\gamma \rightarrow - \infty) = \frac{\pi^2 (n-1)^2}{2 m L^2}, \quad 
n = 2,3,4,\dots,\infty.
\end{equation}
\newpage
\begin{figure}[h]
\begin{center}
\vskip-1.5cm
\includegraphics[width=0.5\textwidth,angle=-90]{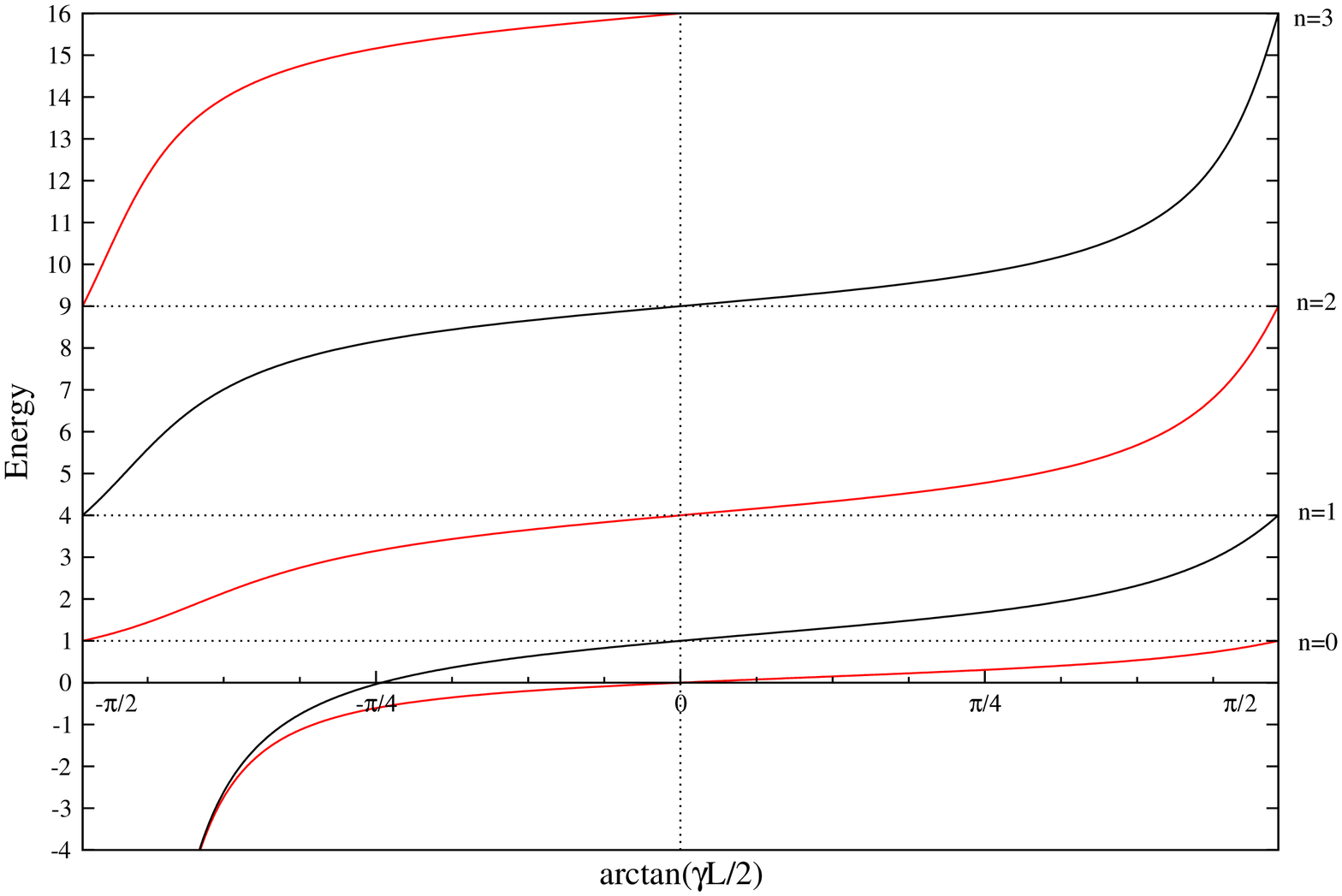} \\ \vskip0.5cm
\includegraphics[width=0.6\textwidth,angle=-90]{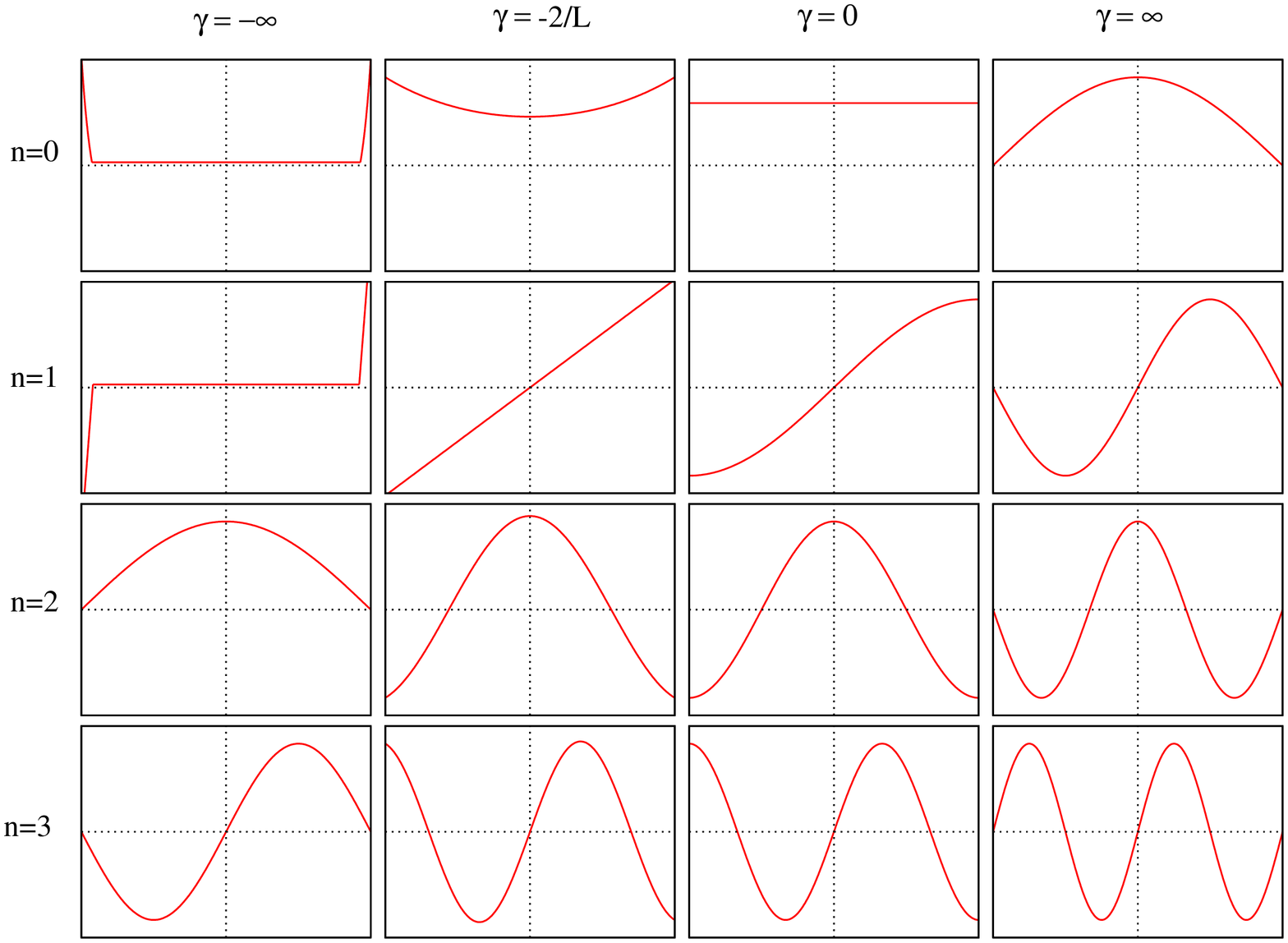}
\end{center}
\caption{\it Top panel: Energy spectrum of a particle in a 1-d box as a function
of the self-adjoint extension parameter $\gamma$. The $x$-value represents
$\arctan(\gamma L/2) \in [- \tfrac{\pi}{2},\tfrac{\pi}{2}]$, which corresponds 
to $\gamma \in [-\infty,\infty]$. The $y$-value represents the energies $E_n$ 
(with $n = 0, 1, 2, 3, 4$) in units of $\pi^2/(2 m L^2)$. Bottom panel: The 
wave functions $\Psi_n(x)$, $x \in [- \tfrac{L}{2},\tfrac{L}{2}]$, (with
$n = 0, 1, 2, 3$), for $\gamma = - \infty, - \tfrac{2}{L}, 0, \infty$. The 
sharp peaks in the $n=0$ and $n=1$ states at $\gamma = - \infty$ represent 
$\delta$-function-type wave functions of negative infinite energy localized at 
the boundaries. Except for these states, the energies and wave functions at 
$\gamma = - \infty$ are the same as those at $\gamma = \infty$.} 
\end{figure}
\newpage
The existence of negative energy states is somewhat counter-intuitive. In 
particular, since the boundary conditions are perfectly reflecting, one may
expect that the particle cannot bind to the wall. However, quantum mechanics
does indeed allow the existence of walls which are perfectly reflecting for
positive energy states and, at the same time, ``sticky'' for negative 
energy states. Since negative energy states have 
$E = \langle p^2/2 m\rangle < 0$, they seem to violate the Heisenberg 
uncertainty relation $\Delta x \Delta p \geq \frac{1}{2}$. Obviously, any state 
confined to the box has $\Delta x \leq L/2$. Hence, the Heisenberg uncertainty 
relation seems to suggest that $\Delta p \geq 1/L$, which would be in conflict 
with $\langle p^2 \rangle < 0$. To resolve this puzzle, one must realize that 
the Heisenberg uncertainty relation was derived for an infinite volume, and thus
needs to be reconsidered in a finite box with perfectly reflecting boundary
conditions.

\subsection{Uncertainty Relation for a 1-d Box with Perfectly Reflecting
Boundary Conditions}

Let us derive a variant of the Heisenberg uncertainty relation, taking into
account the boundary condition eq.(\ref{bc}). We follow the standard procedure
by constructing the non-negative integral
\begin{equation}
I = \int_{-L/2}^{L/2} dx \ |\p_x \Psi(x) + \alpha x \Psi(x) + \beta \Psi(x)|^2 
\geq 0.
\end{equation}
Here $\alpha \in \R$ and $\beta = \beta_r + i \beta_i \in \C$ are parameters to
be varied in order to minimize $I$ and thus derive the most stringent 
inequality. Using partial integration as well as the boundary condition 
eq.(\ref{bc}), it is straightforward to obtain
\begin{equation}
\label{I}
I = \langle p^2 \rangle - \alpha + \beta_r^2 + \beta_i^2 +
\alpha^2 \langle x^2 \rangle + 2 \alpha \beta_r \langle x \rangle + 
2 \beta_i \overline p + \alpha a - b + \beta_r c,
\end{equation}
where we have introduced
\begin{eqnarray}
&&a = \frac{L}{2} [\rho(L/2) + \rho(-L/2)], \nonumber \\
&&b = \gamma [\rho(L/2 + \rho(-L/2)], \nonumber \\
&&c = \rho(L/2) - \rho(-L/2), \quad \rho(\pm L/2) = |\Psi(\pm L/2)|^2.
\end{eqnarray}
We have also defined
\begin{equation}
\overline p = \Re \int_{-L/2}^{L/2} dx \ \Psi(x)^* (- i \p_x) \Psi(x).
\end{equation}
If $p = - i \p_x$ were a self-adjoint operator, $\overline p$ would simply be
the momentum expectation value. However, since $p$ is not self-adjoint in this
case, the corresponding expectation value is in general complex. In particular,
$\overline p$ is not the expectation value of any observable associated with a
self-adjoint operator.

By varying $\alpha$, $\beta_r$, and $\beta_i$ in order to minimize $I$, one 
obtains
\begin{eqnarray}
&&\frac{\p I}{\p \alpha} = - 1 + 2 \alpha \langle x^2 \rangle + 
2 \beta_r \langle x \rangle + a = 0, \nonumber \\
&&\frac{\p I}{\p \beta_r} = 2 \beta_r + 2 \alpha \langle x \rangle + c = 0 \
\Rightarrow \ \beta_r = - \alpha \langle x \rangle - \frac{c}{2}, \nonumber \\
&&\frac{\p I}{\p \beta_i} = 2 \beta_i + 2 \overline p = 0 \
\Rightarrow \ \beta_i = - \overline p,
\end{eqnarray}
which implies
\begin{equation}
\alpha = \frac{1 + c \langle x \rangle - a}{2 (\Delta x)^2}, \quad
(\Delta x)^2 = \langle x^2 \rangle - \langle x \rangle^2 = 
\langle (x - \langle x \rangle)^2 \rangle. 
\end{equation}
Inserting these results for $\alpha$, $\beta_r$, and $\beta_i$ back into the
expression for $I$, eq.(\ref{I}), one obtains
\begin{equation}
I = \langle p^2 \rangle - {\overline p}^2 - 
\left(\frac{1 + c \langle x \rangle - a}{2 \Delta x}\right)^2 - b - 
\frac{c^2}{4} \geq 0,
\end{equation}
which implies the generalized uncertainty relation
\begin{equation}
\label{uncertain}
\langle p^2 \rangle \geq {\overline p}^2 + 
\left(\frac{1 + c \langle x \rangle - a}{2 \Delta x}\right)^2 + b + 
\frac{c^2}{4}.
\end{equation}
In the infinite volume limit, the wave function vanishes at infinity, such that 
$a = b = c = 0$. Furthermore, the momentum operator would then be self-adjoint
with $\langle p \rangle = \overline p$. Hence, in the infinite volume limit, the
inequality (\ref{uncertain}) reduces to the standard Heisenberg uncertainty
relation
\begin{equation}
\langle p^2 \rangle \geq \langle p \rangle^2 + 
\left(\frac{1}{2 \Delta x}\right)^2 \ \Rightarrow \
\Delta x \Delta p \geq \frac{1}{2}, \quad 
(\Delta p)^2 = \langle p^2 \rangle - \langle p \rangle^2 =
\langle (p - \langle p \rangle)^2 \rangle.
\end{equation}

Interestingly, thanks to time-reversal invariance, the eigenfunctions of the
particle in the 1-d box are real-valued and thus $\overline p = 0$. Since
$H = p^2/2m$, for energy eigenstates the generalized uncertainty relation
takes the form
\begin{equation}
2 m E_n = \langle p^2 \rangle \geq 
\left(\frac{1 + c \langle x \rangle - a}{2 \Delta x}\right)^2 + b + 
\frac{c^2}{4}.
\end{equation}
The zero-energy eigenstate $\Psi_0(x) = \sqrt{1/L}$ for $\gamma = 0$ has
$a = 1$, $b = c = 0$, such that the uncertainty relation reduces to
$2 m E_0 \geq 0$, which is indeed satisfied as an equality. This means that this
state represents a minimal uncertainty wave packet. The zero-energy
state $\Psi_1(x) = \sqrt{12/L^3} \ x$ for $\gamma = - 2/L$, on the other hand, 
has $a = 3$, $b = - 12/L^2$, $c = 0$, as well as $\Delta x = \sqrt{3/20} \ L$, 
such that the inequality reduces to $2 m E_1 \geq - 16/(3 L^2)$, which is again 
satisfied, but not saturated as an equality. Hence, in this case, the state
does not represent a minimal uncertainty wave packet.

\subsection{Constructing a Wall with $\gamma < \infty$}

It is interesting to ask whether perfectly reflecting walls with 
$\gamma < \infty$ are just a mathematical curiosity or whether they can also be
constructed physically. Of course, it is clear that, independent of the value of
$\gamma$, a perfectly reflecting wall is always a mathematical idealization.
Any real wall will eventually be penetrable if it is hit by a sufficiently 
energetic particle. In this context it may be interesting to mention the MIT 
bag model \cite{Cho74,Cho74a,Has78}, in which quarks are confined by 
restricting them to a finite region
of space with perfectly reflecting boundary conditions. While this model is
quite successful in modeling the confinement of quarks, in the fundamental QCD
theory confinement does not arise through perfectly reflecting walls but through
confining strings. When a dynamical QCD string connects a very energetic 
quark-anti-quark pair, the string can break by the dynamical creation of further
quark-anti-quark pairs. Since perfectly reflecting walls do not even exist in
the confining theory of the strong force, they certainly do not exist in 
condensed matter either. However, when low-energy particles hit a very high 
energy barrier, it acts effectively like a reflecting wall. In this sense, the 
self-adjoint extension parameter $\gamma$ can be viewed as a low-energy 
parameter that characterizes the reflection properties of a very high energy 
barrier.

Since $\gamma$ appears as a natural mathematical parameter characterizing a
perfectly reflecting wall, there is no reason to expect that it cannot be
physically realized. To show this explicitly, we now construct a wall with an 
arbitrary value of $\gamma$ as a limit of square-well potentials.\footnote{We 
like to thank F.\ Niedermayer for suggesting this construction.} 
\begin{figure}[h]
\begin{center}
\vskip-1cm
\includegraphics[width=0.65\textwidth,angle=-90]{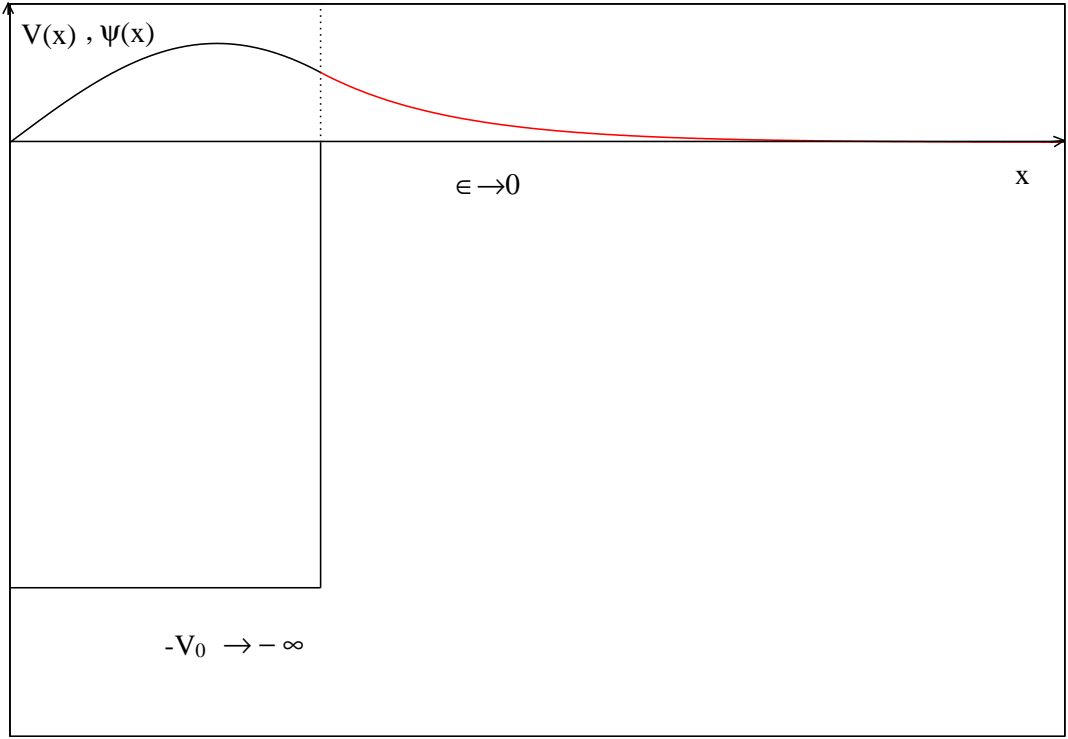}
\end{center}
\caption{\it A deep and narrow square-well potential $V(x)$ with depth
$- V_0 \rightarrow - \infty$ and width $\epsilon \rightarrow 0$ mimics a 
boundary with non-standard self-adjoint extension parameter $\gamma < \infty$. 
The wave function $\Psi(x)$ and its first derivative are continuous at 
$x = \epsilon$. In the example shown here, $\gamma < 0$, such that a bound state
is localized at the boundary.}
\end{figure}
Let us consider the potential illustrated in figure 2, which consists of a 
perfectly reflecting wall at $x = 0$ with the standard textbook value 
$\gamma = \infty$, and a very narrow and very deep square-well potential of 
size $\epsilon > 0$ and depth $- V_0 < 0$ next to it. The wave function then 
takes the form
\begin{equation}
\Psi(x) = A \sin(q x), \quad 0 \leq x \leq \epsilon, \quad 
E = \frac{q^2}{2 m} - V_0.
\end{equation}
For $x \geq \epsilon$, the wave function is determined by enforcing 
continuity of the wave function itself and its derivative at $x = \epsilon$.
When $\epsilon \rightarrow 0$, we can determine the resulting value of $\gamma$ 
from
\begin{equation}
- \gamma \Psi(\epsilon) + \p_x \Psi(\epsilon) = 0 \ \Rightarrow \
\gamma = q \cot(q \epsilon).
\end{equation}
In order to keep $\gamma$ fixed as $\epsilon \rightarrow 0$, we must hence let 
$q$ go to infinity such that
\begin{equation}
q = \frac{\pi}{2 \epsilon} - \frac{2}{\pi} \gamma.
\end{equation}
For all states of finite energy $E$, this is achieved by sending 
$V_0 = q^2/2m \rightarrow \infty$, in such a way that
\begin{equation}
V_0 = \frac{1}{2m} \left(\frac{\pi}{2 \epsilon} - \frac{2}{\pi} \gamma\right)^2.
\end{equation}
It goes without saying that a perfectly reflecting wall with an arbitrary value
of $\gamma$ can also be constructed in many other ways, for example, by using an
attractive $\delta$-function potential next to the wall.

\subsection{Experimental Determination of $\gamma$}

It is natural to ask how one can determine the value of $\gamma$ for some
perfectly reflecting wall that can be investigated experimentally. For example, 
for a planar homogeneous perfectly reflecting wall, the material-specific 
parameter $\gamma$ can be determined from the scattering phase shift 
$\delta(k)$ of an incident plane wave that propagates perpendicular to the 
surface. Assuming that the incident wave propagates in the negative 
$x$-direction in the region $x > 0$, and scatters off a perfectly reflecting 
wall at $x = 0$, the wave function takes the form
\begin{equation}
\Psi(x) = \exp(- i k x) + R \exp(i k x), \quad x \geq 0.
\end{equation}
Imposing the boundary condition
\begin{equation}
\label{wallboundary}
- \gamma \Psi(0) + \p_x \Psi(0) = 0, 
\end{equation}
one obtains 
\begin{equation}
- \gamma (1 + R) - i k (1 - R) = 0 \ \Rightarrow \ R = \exp(i \delta(k)) =
- \frac{\gamma + i k}{\gamma - ik}.
\end{equation}
Hence, by measuring the phase shift
\begin{equation}
\delta(k) = 2 \arctan(k/\gamma) + \pi,
\end{equation}
one can determine the material-specific self-adjoint extension parameter 
$\gamma$.

\section{Uncertainty Relation for a Quantum Dot with General Perfectly
Reflecting Boundary Conditions}

In this section we consider an arbitrarily shaped $d$-dimensional region 
$\Omega$ with general perfectly reflecting walls at the boundary $\p \Omega$.
This may be viewed as a model for a quantum dot, in which electrons are 
confined inside a finite region of space. As illustrated in figure 3, $\Omega$
may have multiple disconnected boundaries. For a mathematical exposition 
of boundary value problems for operator differential equations we refer to 
\cite{Gor91}.
\begin{figure}[h]
\begin{center}
\includegraphics[width=0.65\textwidth,angle=-90]{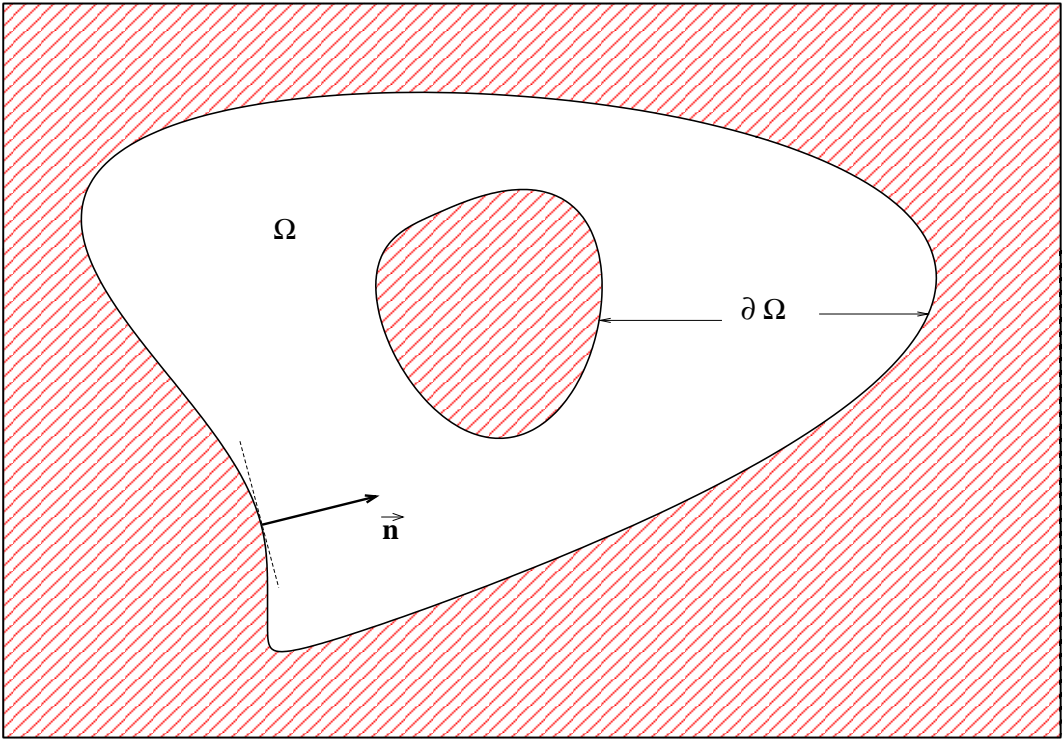}
\end{center}
\caption{\it A region $\Omega$ of a quantum dot with two disconnected
boundaries $\p \Omega$. The shaded region is energetically forbidden. The 
vector $\vec n$ is normal to the boundary.}
\end{figure}
Let us consider the Hamiltonian 
\begin{equation}
H = \frac{{\vec p \,}^2}{2 m} + V(\vec x) = - \frac{1}{2 m} \Delta + V(\vec x),
\end{equation}
where $V(\vec x)$ is a non-singular potential. In $d$ dimensions, the 
continuity equation for local probability conservation takes the form
\begin{equation}
\p_t \rho(\vec x,t) + \vec \nabla \cdot \vec j(\vec x,t) = 0,
\end{equation}
with the probability density and current density given by
\begin{equation}
\rho(\vec x,t) = |\Psi(\vec x,t)|^2, \quad
\vec j(\vec x,t) = \frac{1}{2 m i}
\left[\Psi(\vec x,t)^* \vec \nabla \Psi(\vec x,t) -
\vec \nabla \Psi(\vec x,t)^* \Psi(\vec x,t)\right].
\end{equation}
Again, in the following discussion the time-dependence is not essential and we 
simplify the notation to $\Psi(\vec x)$.

\subsection{Spatial Boundary Conditions}

As in the 1-d case, we demand that no probability may leak outside the region
$\Omega$. This is ensured by requiring that the component of the probability
current density normal to the surface vanishes, i.e.
\begin{equation}
\vec n(\vec x) \cdot \vec j(\vec x) = 0, \quad \vec x \in \p \Omega,
\end{equation}
where $\vec n(\vec x)$ is the unit-vector normal to the surface $\p \Omega$ at 
the point $\vec x$. The most general local boundary condition that ensures this 
takes the form
\begin{equation}
\label{bcdot}
\gamma(\vec x) \Psi(\vec x) + \vec n(\vec x) \cdot \vec \nabla \Psi(\vec x) = 0,
\quad x \in \p \Omega,
\end{equation}
and one then indeed obtains
\begin{equation}
\vec n(\vec x) \cdot \vec j(\vec x) = \frac{1}{2 m i}
[- \Psi(\vec x)^* \gamma(\vec x) \Psi(\vec x,t) + 
\gamma(\vec x)^* \Psi(\vec x,t)^* \Psi(\vec x,t)] = 0, \quad 
\vec x \in \p \Omega,
\end{equation}
which again implies $\gamma(\vec x) \in \R$. In general $\gamma(\vec x)$ will
depend on the position $\vec x \in \p \Omega$ on the boundary. In a real system
such as a quantum dot, $\gamma(\vec x)$ is a material-specific parameter, to be
determined experimentally. 

\subsection{Self-Adjointness of the Hamiltonian}

Let us again convince ourselves that the Hamiltonian endowed with the boundary
condition eq.(\ref{bcdot}) is indeed self-adjoint. For this purpose, we consider
\begin{eqnarray}
\langle\chi|H|\Psi\rangle&=&
\int_\Omega d^dx \ \chi(\vec x)^* 
\left[- \frac{1}{2 m} \Delta + V(\vec x)\right] \Psi(\vec x) \nonumber \\
&=&\int_\Omega d^dx \ 
\left[\frac{1}{2 m} \vec \nabla \chi(\vec x)^* \cdot \vec \nabla \Psi(\vec x) +
\chi(\vec x)^* V(\vec x) \Psi(\vec x) \right] \nonumber \\
&-&\frac{1}{2 m}
\int_{\p \Omega} d\vec n \cdot \chi(\vec x)^* \vec \nabla \Psi(\vec x)
\nonumber \\
&=&\int_\Omega d^dx \ \left\{\left[- \frac{1}{2 m} \Delta + V(\vec x)\right]
\chi(\vec x)^* \right\} \Psi(\vec x) \nonumber \\
&+&\frac{1}{2 m} \int_{\p \Omega} d\vec n \cdot
\left[\vec \nabla \chi(\vec x)^* \Psi(\vec x) - 
\chi(\vec x)^* \vec \nabla \Psi(\vec x)\right] \nonumber \\
&=&\langle\Psi|H|\chi\rangle^* + \frac{1}{2 m} \int_{\p \Omega} d\vec n \cdot
\left[\vec \nabla \chi(\vec x)^* \Psi(\vec x) - 
\chi(\vec x)^* \vec \nabla \Psi(\vec x)\right].
\end{eqnarray}
Hence, the Hamiltonian is Hermitean if
\begin{equation}
\label{symmetricHdot}
\int_{\p \Omega} d\vec n \cdot
\left[\vec \nabla \chi(\vec x)^* \Psi(\vec x) - 
\chi(\vec x)^* \vec \nabla \Psi(\vec x)\right] = 0.
\end{equation}
Using the boundary condition eq.(\ref{bcdot}), the integral in 
eq.(\ref{symmetricHdot}) reduces to
\begin{equation}
\int_{\p \Omega} d^{d-1}x \left[\vec n(\vec x) \cdot \vec \nabla \chi(\vec x)^* + 
\gamma(\vec x) \chi(\vec x)^*\right] \Psi(\vec x) = 0.
\end{equation}
Since $\Psi(\vec x)$ itself can take arbitrary values at the boundary, the
Hermiticity of $H$ requires that
\begin{equation}
\vec n(\vec x) \cdot \vec \nabla \chi(\vec x) + \gamma(\vec x)^* \chi(\vec x) = 
0.
\end{equation}
For $\gamma(\vec x) \in \R$, this is again the boundary condition of 
eq.(\ref{bcdot}), which ensures that $D(H^\dagger) = D(H)$, such that $H$ is
indeed self-adjoint. In complete analogy to the 1-d case, it is easy to convince
oneself that the momentum operator $\vec p = - i \vec \nabla$ is not even
Hermitean in the domain $D(H)$ of the Hamiltonian, which contains those 
twice-differentiable and square-integrable wave functions that obey the
boundary condition eq.(\ref{bcdot}).

\subsection{Generalized Uncertainty Relation}

Mathematical investigations of generalized uncertainty relations can be found, 
for example, in \cite{Fef83,Esc10}. In this subsection, we derive a generalized 
uncertainty relation for an arbitrarily shaped quantum dot in $d$ dimensions, 
by considering the non-negative integral
\begin{equation}
\label{Idot1}
I = \int_\Omega d^dx \ |\vec \nabla \Psi(\vec x) + \alpha \vec x \Psi(\vec x) + 
\vec \beta \Psi(\vec x)|^2 \geq 0.
\end{equation}
Again, $\alpha \in \R$ and $\vec \beta = \vec \beta_r + i \vec \beta_i \in \C^d$
are parameters to be varied in order to minimize $I$. Using the boundary
condition eq.(\ref{bcdot}) after performing some partial integrations, one finds
\begin{equation}
\label{Idot2}
I = \langle {\vec p \,}^2 \rangle - \alpha d + \vec \beta_r^2 + \vec \beta_i^2 +
\alpha^2 \langle {\vec x \,}^2 \rangle + 
2 \alpha \vec \beta_r \cdot \langle \vec x \rangle + 
2 \vec \beta_i \cdot \overline{\vec p} + 
\alpha \langle \vec n \cdot \vec x \rangle - 
\langle \gamma \rangle + 
\vec \beta_r \cdot \langle \vec n \rangle.
\end{equation}
Here we have defined
\begin{eqnarray}
&&\langle \vec n \cdot \vec x \rangle =
\int_{\p \Omega} d\vec n \cdot \vec x \rho(\vec x), \nonumber \\
&&\langle \gamma \rangle = 
\int_{\p \Omega} d^{d-1}x \ \gamma(\vec x) \rho(\vec x), \nonumber \\
&&\langle \vec n \rangle =
\int_{\p \Omega} d\vec n \ \rho(\vec x), \quad 
\rho(\vec x) = |\Psi(\vec x)|^2.
\end{eqnarray}
In analogy to the 1-d case, we have also introduced
\begin{equation}
\overline{\vec p} = \Re \int_\Omega d^dx \ 
\Psi(\vec x)^* (- i \vec \nabla) \Psi(\vec x).
\end{equation}
If $\vec p = - i \vec \nabla$ had been a self-adjoint operator, 
$\overline{\vec p}$ would be the momentum expectation value. However, in this
case, $\overline{\vec p}$ is again not the expectation value of any observable 
physical quantity.

By varying $\alpha$, $\vec \beta_r$, and $\vec \beta_i$, one obtains
\begin{eqnarray}
&&\frac{\p I}{\p \alpha} = - d + 2 \alpha \langle {\vec x \,}^2 \rangle + 
2 \vec \beta_r \cdot \langle \vec x \rangle + 
\langle \vec n \cdot \vec x \rangle = 0, \nonumber \\
&&\frac{\p I}{\p \beta_r^j} = 2 \beta_r^j + 2 \alpha \langle x^j \rangle + 
\langle n^j \rangle =
0 \ \Rightarrow \ 
\vec \beta_r = - \alpha \langle \vec x \rangle - 
\frac{\langle \vec n \rangle}{2}, \nonumber \\
&&\frac{\p I}{\p \beta_i^j} = 2 \beta_i^j + 2 \overline{p^j} = 0 \
\Rightarrow \ \vec \beta_i = - \overline{\vec p},
\end{eqnarray}
which then implies
\begin{equation}
\alpha = \frac{d + 
\langle \vec n \rangle \cdot \langle \vec x \rangle - 
\langle \vec n \cdot \vec x \rangle}{2 (\Delta x)^2}, 
\quad (\Delta x)^2 = \langle {\vec x \,}^2 \rangle - \langle \vec x \rangle^2 = 
\langle (\vec x - \langle \vec x \rangle)^2 \rangle. 
\end{equation}
Again, by inserting the results for $\alpha$, $\vec \beta_r$, and 
$\vec \beta_i$ back into eq.(\ref{Idot2}), one obtains
\begin{equation}
I = \langle {\vec p \,}^2 \rangle - {\overline{\vec p} \,}^2 - 
\left(\frac{d + 
\langle \vec n \rangle \cdot \langle \vec x \rangle - 
\langle \vec n \cdot \vec x \rangle}{2 \Delta x}\right)^2 - 
\langle \gamma \rangle - 
\frac{\langle \vec n \rangle^2}{4} \geq 0,
\end{equation}
which finally implies the generalized uncertainty relation
\begin{equation}
\label{uncertaindot}
\langle {\vec p \,}^2 \rangle \geq {\overline{\vec p} \,}^2 + 
\left(\frac{d + 
\langle \vec n \rangle \cdot \langle \vec x \rangle - 
\langle \vec n \cdot \vec x \rangle}{2 \Delta x}\right)^2 + 
\langle \gamma \rangle +
\frac{\langle \vec n \rangle^2}{4}.
\end{equation}

It is not obvious that eq.(\ref{uncertaindot}) is invariant under a spatial
translation $\vec x \rightarrow {\vec x}\,' = \vec x + \vec d$. While 
$\langle {\vec p \,}^2 \rangle$, $\overline{\vec p}$, $\Delta x$, 
$\langle \gamma \rangle$, and $\langle \vec n \rangle$ 
are translation invariant by construction, $\langle \vec x \rangle$ and
$\langle \vec n \cdot \vec x \rangle$ are not. It is thus reassuring 
to realize that the combination
\begin{equation}
\langle \vec n \rangle \cdot \langle {\vec x}\,' \rangle - 
\langle \vec n \cdot {\vec x}\,' \rangle =
\langle \vec n \rangle \cdot (\langle \vec x \rangle + \vec d) - 
\langle \vec n \cdot \vec x \rangle - 
\langle \vec n \cdot \vec d \rangle =
\langle \vec n \rangle \cdot \langle \vec x \rangle - 
\langle \vec n \cdot \vec x \rangle,
\end{equation}
that enters the generalized uncertainty relation is indeed translation
invariant.

All quantities entering the uncertainty relation, except $\overline{\vec p}$,
are directly related to physically observable quantities. However, since the
momentum is not a self-adjoint operator and hence not observable in a finite
volume, despite the fact that it is mathematically completely well-defined,
the quantity $\overline{\vec p}$ seems not to be physically measurable. In that
case, the question arises how the generalized uncertainty relation should be
interpreted physically. Since momentum cannot even be measured inside a quantum 
dot, $(\Delta p)^2 = \langle {\vec p \,}^2 \rangle - {\overline{\vec p} \,}^2$ 
can obviously not be interpreted as the uncertainty of momentum. Still, 
$\langle {\vec p \,}^2 \rangle$ determines the energy of an eigenstate of the
free particle Hamiltonian with $V(\vec x) = 0$. Since time-reversal invariance 
guarantees that energy eigenstates have real-valued wave functions, for those 
states we know that $\overline{\vec p} = 0$. In that case, the generalized 
uncertainty relation reduces to
\begin{equation}
2 m E_n = \langle {\vec p \,}^2 \rangle \geq \left(\frac{d + 
\langle \vec n \rangle \cdot \langle \vec x \rangle - 
\langle \vec n \cdot \vec x \rangle}{2 \Delta x}\right)^2 + 
\langle \gamma \rangle +
\frac{\langle \vec n \rangle^2}{4},
\end{equation}
which indeed contains measurable physical quantities only.

\subsection{General Uncertainty Relation for non-Hermitean Operators}

For two general self-adjoint operators $A$ and $B$, the uncertainty relation 
is sometimes quoted as 
$\Delta A \Delta B \geq \frac{1}{2} |\langle [A,B] \rangle|$. However, a more 
stringent form of the inequality is given by
\begin{equation}
\Delta A \Delta B \geq |\langle \widetilde A \widetilde B \rangle| =
|\langle A B \rangle - \langle A \rangle \langle B \rangle|, \quad
\widetilde A = A - \langle A \rangle, \quad
\widetilde B = B - \langle B \rangle.
\end{equation}
In the case discussed before, the momentum operator is not even Hermitean.
Hence, the question arises, how the uncertainty relation looks like for a 
general pair of non-Hermitean operators $A \neq A^\dagger$ and 
$B \neq B^\dagger$. In that case, it is natural to define
\begin{equation}
(\Delta A)^2 = \langle \widetilde A \Psi|\widetilde A \Psi \rangle =
\langle \widetilde A^\dagger \widetilde A \rangle =
\langle A^\dagger A \rangle - \langle A^\dagger \rangle \langle A \rangle =
\langle A^\dagger A \rangle - |\langle A \rangle|^2 \geq 0.
\end{equation}
Here we have used $\langle A^\dagger \rangle = \langle A \rangle^*$. Using the 
Schwarz inequality, we then obtain
\begin{eqnarray}
(\Delta A)^2 (\Delta B)^2&=& 
\langle \widetilde A \Psi|\widetilde A \Psi \rangle 
\langle \widetilde B \Psi|\widetilde B \Psi \rangle \nonumber \\
&\geq&|\langle \widetilde A \Psi|\widetilde B \Psi \rangle|^2 = 
|\langle \widetilde A^\dagger\widetilde B \rangle|^2 =
|\langle A^\dagger B \rangle - \langle A^\dagger \rangle \langle B \rangle|^2,
\end{eqnarray}
such that
\begin{equation}
\label{generaluncertainty}
\Delta A \Delta B \geq 
|\langle A^\dagger B \rangle - \langle A^\dagger \rangle \langle B \rangle|.
\end{equation}

Let us compare the general uncertainty relation (\ref{generaluncertainty})
for non-Hermitean operators with the generalized uncertainty relation 
(\ref{uncertaindot}) that we obtained for a particle confined to the region
$\Omega$. In that case, we can identify $A$ with the self-adjoint operator 
$\vec x$ and $B$ with the non-Hermitean operator $\vec p = - i \vec \nabla$.
The uncertainty of the momentum is then given by
\begin{equation}
(\Delta p)^2 = 
\langle {\vec p \,}^\dagger \cdot \vec p \,\rangle - 
|\langle \vec p \,\rangle|^2.
\end{equation}
Since ${\vec p \,}^\dagger \neq \vec p$, in this case, 
${\vec p \,}^\dagger \cdot \vec p \neq {\vec p \,}^2$. In particular, we obtain
\begin{eqnarray}
\langle {\vec p \,}^\dagger \cdot \vec p \,\rangle&=&
\langle \vec p \, \Psi|\vec p \, \Psi \rangle =
\int_\Omega d^dx \ \vec \nabla \Psi(\vec x)^* \cdot \vec \nabla \Psi(\vec x) 
\nonumber \\
&=&- \int_\Omega d^dx \ \Psi(\vec x)^* \Delta \Psi(\vec x) +
\int_{\p \Omega} d\vec n \cdot \Psi(\vec x)^* \vec \nabla \Psi(\vec x) 
\nonumber \\
&=&\langle {\vec p \,}^2 \rangle - 
\int_{\p \Omega} d^{d-1}x \ \gamma(\vec x) \Psi(\vec x)^* \Psi(\vec x) =
\langle {\vec p \,}^2 \rangle - \langle \gamma \rangle.
\end{eqnarray}
Furthermore, decomposing $\langle \vec p \,\rangle$ into real and imaginary 
parts, we find
\begin{equation}
\langle \vec p \, \rangle = \Re \int_\Omega d^dx \ 
\Psi(\vec x)^* (- i \vec \nabla) \Psi(\vec x) + 
i \Im \int_\Omega d^dx \ \Psi(\vec x)^* (- i \vec \nabla) \Psi(\vec x) =
\overline{\vec p} - \frac{i}{2} \langle \vec n \rangle.
\end{equation}
Here we have used
\begin{eqnarray}
\Im \int_\Omega d^dx \ \Psi(\vec x)^* (- i \vec \nabla) \Psi(\vec x)&=&
\frac{1}{2 i} \int_\Omega d^dx 
\left[\Psi(\vec x)^* (- i \vec \nabla) \Psi(\vec x) -
\Psi(\vec x) (i \vec \nabla) \Psi(\vec x)^*\right] \nonumber \\
&=&- \frac{1}{2} \int_{\p \Omega} d\vec n \ \Psi(\vec x)^* \Psi(\vec x) =
- \frac{1}{2} \langle \vec n \rangle,
\end{eqnarray}
and hence we obtain
\begin{equation}
(\Delta p)^2 = \langle {\vec p \,}^2 \rangle - \langle \gamma \rangle
- {\overline{\vec p} \,}^2 - \frac{\langle \vec n \rangle^2}{4}.
\end{equation}
The generalized uncertainty relation (\ref{uncertaindot}) can thus be rewritten 
as 
\begin{equation}
\Delta x \Delta p \geq \frac{1}{2} 
|d + \langle \vec n \rangle \cdot \langle \vec x \rangle - 
\langle \vec n \cdot \vec x \rangle|.
\end{equation}
On the other hand, in this case the general uncertainty relation for 
non-Hermitean operators (\ref{generaluncertainty}) takes the form
\begin{equation}
\Delta x \Delta p \geq
|\langle \vec x \cdot \vec p \,\rangle - \langle \vec x \rangle \cdot 
\langle \vec p \,\rangle|.
\end{equation}
Introducing 
$\overline{\vec x \cdot \vec p} = \Re \langle \vec x \cdot \vec p \,\rangle$
and using
\begin{eqnarray}
\Im \langle \vec x \cdot \vec p \,\rangle&=&
\frac{1}{2i} \int_\Omega d^dx 
\left[\Psi(\vec x)^* \vec x \cdot (- i \vec \nabla) \Psi(\vec x) -
\Psi(\vec x) \vec x \cdot (i \vec \nabla) \Psi(\vec x)^*\right] \nonumber \\
&=&\frac{1}{2} 
\int_\Omega d^dx \ \Psi(\vec x)^* \vec \nabla \cdot \vec x \, \Psi(\vec x) -
\frac{1}{2} 
\int_{\p \Omega} d\vec n \cdot \vec x \, \Psi(\vec x)^* \Psi(\vec x) \nonumber \\
&=&\frac{1}{2}(d - \langle \vec n \cdot \vec x \rangle),
\end{eqnarray}
one obtains
\begin{equation}
\langle \vec x \cdot \vec p \,\rangle - \langle \vec x \rangle \cdot 
\langle \vec p \,\rangle = \overline{\vec x \cdot \vec p} -
\langle \vec x \rangle \cdot \overline{\vec p} + 
\frac{i}{2}(d - \langle \vec n \cdot \vec x \rangle + 
\langle \vec n \rangle \cdot \langle \vec x \rangle),
\end{equation}
such that
\begin{equation}
\Delta x \Delta p \geq 
\sqrt{\left(\overline{\vec x \cdot \vec p} -
\langle \vec x \rangle \cdot \overline{\vec p}\right)^2 + 
\frac{1}{4}\left(d - \langle \vec n \cdot \vec x \rangle + 
\langle \vec n \rangle \cdot \langle \vec x \rangle\right)^2}.
\end{equation}
Hence, unless $\overline{\vec x \cdot \vec p} = 
\langle \vec x \rangle \cdot \overline{\vec p}$, the general uncertainty 
relation for non-Hermitean operators (\ref{generaluncertainty}) is more 
stringent than the generalized uncertainty relation (\ref{uncertaindot}). Since
energy eigenstates have a real-valued wave function, for them 
$\overline{\vec x \cdot \vec p} = 0$ and $\overline{\vec p} = 0$, such that
both inequalities are then equivalent.

\subsection{Minimal Uncertainty Wave Packets}

It is interesting to ask which wave functions saturate the generalized 
uncertainty relation (\ref{uncertaindot}), and thus satisfy it as an equality. 
First of all, when $\gamma(\vec x) = 0$ everywhere at the boundary, the constant
wave function $\Psi(\vec x) = 1/\sqrt{V}$, where $V$ is the volume of $\Omega$, 
is an energy eigenstate of zero energy. In that case, we obtain 
$\overline{\vec p} = 0$, $\langle \gamma \rangle = 0$, as well as
\begin{eqnarray}
&&\langle \vec n \cdot \vec x \rangle = 
\frac{1}{V} \int_{\p \Omega} d\vec n \cdot \vec x = 
\frac{1}{V} \int_\Omega d^dx \ \vec \nabla \cdot \vec x = d, \nonumber \\
&&\langle \vec n \rangle = \frac{1}{V} \int_{\p \Omega} d\vec n =
\frac{1}{V} \int_\Omega d^dx \ \vec \nabla 1 = 0,
\end{eqnarray}
such that the inequality (\ref{uncertaindot}) then reduces to 
$2 m E = \langle {\vec p \,}^2 \rangle \geq 0$. Hence, the zero-energy state
indeed saturates the inequality. In this sense, it can be viewed as a state of
minimal uncertainty. Of course, one should not forget that, since in this case
momentum is not even a physical observable, $\langle {\vec p \,}^2 \rangle$ 
cannot be interpreted as the uncertainty of a momentum measurement.

Are there other minimal uncertainty wave packets beyond the zero-energy state
with a constant wave function that exists for $\gamma(\vec x) = 0$? It is clear
by construction, that the inequality (\ref{uncertaindot}) can be saturated only
if the integrand in eq.(\ref{Idot1}) vanishes, i.e.\ if
\begin{equation}
\label{minimal}
\vec \nabla \Psi(\vec x) + \alpha \vec x \Psi(\vec x) + \vec \beta \Psi(\vec x)
= 0.
\end{equation}
Using the boundary condition eq.(\ref{bcdot}), for points $\vec x \in \p \Omega$
this implies
\begin{equation}
\left(- \gamma(\vec x) + \alpha \vec n(\vec x) \cdot \vec x + 
\vec n(\vec x) \cdot \vec \beta\right)
\Psi(\vec x) = 0 \ \Rightarrow \ 
\gamma(\vec x) = \vec n(\vec x) \cdot (\alpha \vec x + \vec \beta).
\end{equation}
Since $\gamma(\vec x) \in \R$, we must further demand $\vec \beta_i = 
- \overline{\vec p} = 0$. This implies that $\gamma(\vec x)$ must have a very 
peculiar form at the boundary, which will generically not be the case. Thus, in
general, there are no minimal uncertainty wave packets in a finite volume. In 
the bulk eq.(\ref{minimal}) is satisfied just by the standard Gaussian wave 
packet
\begin{equation}
\Psi(\vec x) = 
A \exp\left(- \frac{\alpha}{2}{\vec x \,}^2 - 
\vec \beta \cdot \vec x\right),
\end{equation}
which is known to saturate the Heisenberg uncertainty relation in the infinite
volume. We then also have
\begin{equation}
\alpha = \frac{d + 
\langle \vec n \rangle \cdot \langle \vec x \rangle - 
\langle \vec n \cdot \vec x \rangle}{2 (\Delta x)^2}, \quad
\vec \beta = - \alpha \langle \vec x \rangle - 
\frac{\langle \vec n \rangle}{2}, 
\end{equation}
such that $\gamma(\vec x)$ must satisfy
\begin{equation}
\gamma(\vec x) = 
\vec n(\vec x) \cdot \left(\alpha (\vec x - \langle \vec x \rangle) -
\frac{\langle \vec n \rangle}{2}\right) \ \Rightarrow \
\langle \gamma \rangle = \alpha (\langle \vec n \cdot \vec x \rangle -
\langle \vec n \rangle \cdot \langle \vec x \rangle) -
\frac{\langle \vec n \rangle^2}{2}. 
\end{equation}
Only if $\gamma(\vec x)$ happens to be such that the Gaussian wave 
packet automatically satisfies the boundary condition, it remains a minimal 
uncertainty wave packet in the finite volume.

\subsection{$\gamma$-Dependence of the Energy Spectrum}

In this subsection we assume that $\gamma(\vec x) = \gamma \in \R$ is a constant
independent of the position $\vec x \in \p \Omega$ on the boundary. We then ask
how the energy spectrum changes with $\gamma$. A similar calculation for the 
Dirac operator in a relativistic field theory was performed in \cite{Wip95}. Let
us introduce the energy eigenvalues $E_n$ and the corresponding wave functions 
$\Psi_n(\vec x)$, i.e. 
\begin{equation}
H \Psi_n(\vec x) = \left(- \frac{1}{2m} \Delta + V(\vec x)\right) \Psi_n(\vec x)
= E_n \Psi_n(\vec x).
\end{equation}
Both $E_n$ and $\Psi_n(\vec x)$ depend on $\gamma$ via the boundary condition
\begin{equation}
\gamma \Psi_n(\vec x) + \vec n(\vec x) \cdot \vec \nabla \Psi_n(\vec x) = 0, 
\quad \vec x \in \p \Omega.
\end{equation}
The $\gamma$-dependence of the energy spectrum follows from
\begin{eqnarray}
\label{monotonic}
\p_\gamma E_n&=&\p_\gamma \int_\Omega d^dx \ \Psi_n(\vec x)^* H \Psi_n(\vec x) 
\nonumber \\
&=&\int_\Omega d^dx \ \left[\p_\gamma \Psi_n(\vec x)^* H\Psi_n(\vec x) 
+ \Psi_n(\vec x)^* H \p_\gamma \Psi_n(\vec x)\right] \nonumber \\
&=&\int_\Omega d^dx \ \left[\p_\gamma \Psi_n(\vec x)^* H \Psi_n(\vec x) 
+ H \Psi_n(\vec x)^* \p_\gamma \Psi_n(\vec x)\right] \nonumber \\
&+&\frac{1}{2 m} \int_{\p \Omega} d\vec n \cdot \left[
\vec \nabla \Psi_n(\vec x)^* \p_\gamma \Psi_n(\vec x) -
\Psi_n(\vec x)^* \vec \nabla \p_\gamma \Psi_n(\vec x)\right] \nonumber \\
&=&E_n \ \p_\gamma \int_\Omega d^dx \ \Psi_n(\vec x)^* \Psi_n(\vec x) 
\nonumber \\
&+&\frac{1}{2 m} \int_{\p \Omega} d^{d-1}x \left[
- \gamma \Psi_n(\vec x)^* \p_\gamma \Psi_n(\vec x) +
\Psi_n(\vec x)^* \p_\gamma \left(\gamma \Psi_n(\vec x)\right)\right] 
\nonumber \\
&=&\frac{1}{2 m} \int_{\p \Omega} d^{d-1}x \ \Psi_n(\vec x)^* \Psi_n(\vec x) =
\frac{1}{2 m} \int_{\p \Omega} d^{d-1}x \ \rho_n(\vec x) \geq 0,
\end{eqnarray}
which shows that the spectrum is monotonically rising with $\gamma$. 

Indeed, in the 1-d case, the spectrum illustrated in figure 1 is monotonic in
$\gamma$. In that case, eq.(\ref{monotonic}) reduces to
\begin{equation}
\p_\gamma E_n = \frac{1}{2 m}\left[\rho_n(L/2) + \rho_n(- L/2)\right].
\end{equation}
Let us explicitly verify this equation for the positive energy eigenstates of
even parity
\begin{equation}
\Psi_n(x) = A \cos(k_n x), \quad E_n = \frac{k_n^2}{2 m}, \quad 
\frac{\gamma}{k_n} = \tan \frac{k_n L}{2}.
\end{equation}
In this case, one obtains
\begin{equation}
\p_\gamma E_n = \frac{k_n \p_\gamma k_n}{m} = 
\frac{2 k_n \cos^2(k_n L/2)}{m[k_n L + \sin(k_n L)]},
\end{equation}
as well as
\begin{equation}
\rho_n(L/2) + \rho_n(- L/2) = 2 |A|^2 \cos^2\frac{k_n L}{2}.
\end{equation}
The normalization condition for the wave function implies
\begin{equation}
|A|^2 \int_{-L/2}^{L/2}dx \ \cos^2\frac{k_n L}{2} = 
|A|^2 \frac{1}{2 k_n} \left(k_n L + \sin(k_n L)\right) = 1,
\end{equation}
such that indeed
\begin{equation}
\frac{1}{2 m}\left[\rho_n(L/2) + \rho_n(- L/2)\right] = 
\frac{2 k_n}{m[k_n L + \sin(k_n L)]} \cos^2\frac{k_n L}{2} = \p_\gamma E_n.
\end{equation}
It is straightforward to repeat this check for states of odd parity or negative 
energy.

\subsection{General Boundary Conditions for Heterostructures}

Semiconductor heterostructures such as quantum dots, quantum wires, and 
quantum wells are separated into regions with different effective electron 
masses \cite{Har05}. Until now we have considered quantum dots consisting of a 
single region isolated from an energetically forbidden environment. In this 
subsection, we construct the most general condition at the boundary $\p \Omega$ 
separating two regions, $\Omega_{\I}$ and $\Omega_{\II}$ as illustrated in figure
4, with different effective electron masses $m_{\I}$ and $m_{\II}$. 
\begin{figure}[b]
\begin{center}
\vskip1.3cm
\includegraphics[width=0.65\textwidth,angle=-90]{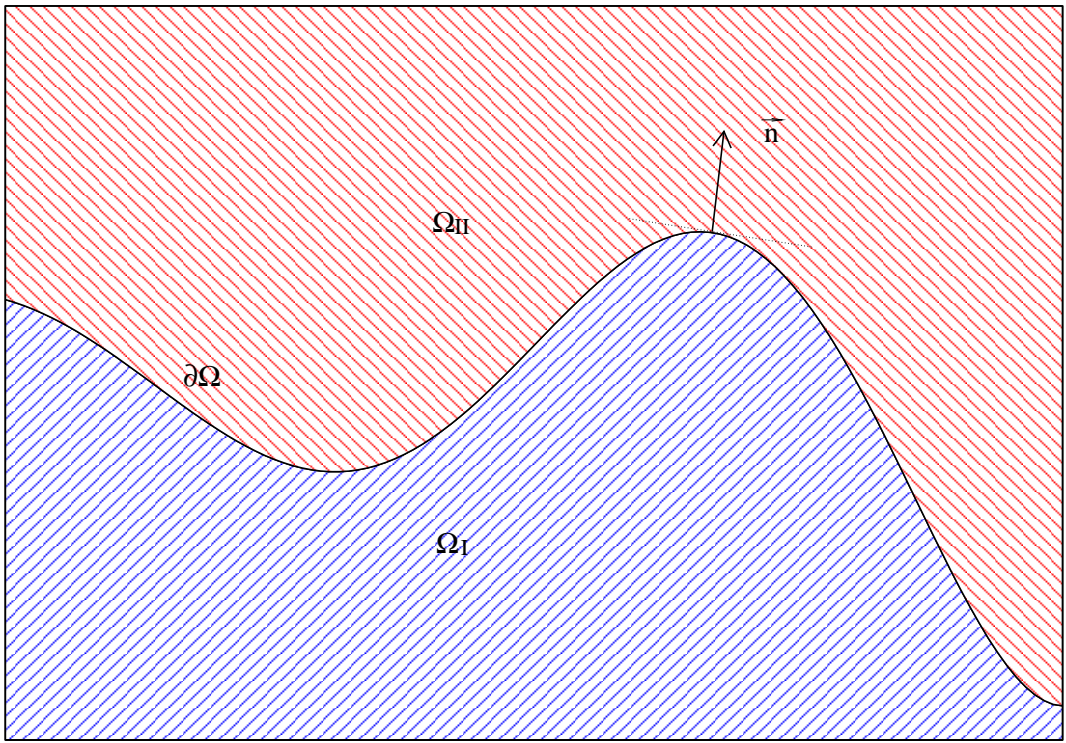}
\end{center}
\caption{\it Two regions $\Omega_{\I}$ and $\Omega_{\II}$ in a quantum 
heterostructure are separated by a boundary $\p \Omega$ with the unit-vector 
$\vec n$ normal to the surface. An electron may have different effective masses 
$m_{\I}$ and $m_{\II}$ in the two regions.}
\end{figure}
In the two regions the Hamiltonian is then given by
\begin{equation}
H_{\I} = \frac{{\vec p \,}^2}{2 m_{\I}} + V_{\I}(\vec x), \quad
H_{\II} = \frac{{\vec p \,}^2}{2 m_{\II}} + V_{\II}(\vec x).
\end{equation}
For points at the boundary $\p \Omega$, the Hermiticity condition takes the form
\begin{eqnarray}
\label{HermiteI,II}
&&\vec n(\vec x) \cdot \frac{1}{2 m_{\I}}
\left[\chi_{\I}(\vec x)^* \vec \nabla \Psi_{\I}(\vec x) -
\vec \nabla \chi_{\I}(\vec x)^* \Psi_{\I}(\vec x)\right] = \nonumber \\
&&\vec n(\vec x) \cdot \frac{1}{2 m_{\II}}
\left[\chi_{\II}(\vec x)^* \vec \nabla \Psi_{\II}(\vec x) - 
\vec \nabla \chi_{\II}(\vec x)^* \Psi_{\II}(\vec x)\right].
\end{eqnarray}
The self-adjointness condition can now be expressed as
\begin{equation}
\left(\begin{array}{c} \Psi_{\I}(\vec x) \\ 
\frac{1}{2 m_{\I}} \vec n(\vec x) \cdot \vec \nabla \Psi_{\I}(\vec x)\end{array}
\right) =
\Gamma(\vec x) \left(\begin{array}{c} \Psi_{\II}(\vec x) \\ 
\frac{1}{2 m_{\II}} \vec n(\vec x) \cdot \vec \nabla \Psi_{\II}(\vec x)
\end{array}\right), \quad \Gamma(\vec x) \in GL(2,\C),
\end{equation}
which turns the Hermiticity condition eq.(\ref{HermiteI,II}) into
\begin{eqnarray}
&&\left[\Gamma_{21}(\vec x) \chi_{\I}(\vec x)^* - \Gamma_{11}(\vec x) 
\frac{1}{2 m_{\I}} \vec n(\vec x) \cdot \vec \nabla \chi_{\I}(\vec x)^* +
\frac{1}{2 m_{\II}} \vec n(\vec x) \cdot \vec \nabla \chi_{\II}(\vec x)^* \right]
\Psi_{\II}(\vec x) + \nonumber \\
&&\left[\Gamma_{22}(\vec x) \chi_{\I}(\vec x)^* - \Gamma_{12}(\vec x) 
\frac{1}{2 m_{\I}} \vec n(\vec x) \cdot \vec \nabla \chi_{\I}(\vec x)^* - 
\chi_{\II}(\vec x)^* \right] \frac{1}{2 m_{\II}} \vec n(\vec x) \cdot 
\vec \nabla \Psi_{\II}(\vec x) = 0. \nonumber \\ \,
\end{eqnarray}
In order to satisfy this relation we must thus demand
\begin{equation}
\left(\begin{array}{c} \chi_{\II}(\vec x) \\ 
\frac{1}{2 m_{\II}} \vec n(\vec x) \cdot \vec \nabla \chi_{\II}(\vec x)
\end{array}\right) =
\left(\begin{array}{cc} \Gamma_{22}(\vec x)^* & - \Gamma_{12}(\vec x)^* \\
- \Gamma_{21}(\vec x)^* & \Gamma_{11}(\vec x)^* \end{array}\right)
\left(\begin{array}{c} \chi_{\I}(\vec x) \\ 
\frac{1}{2 m_{\I}} \vec n(\vec x) \cdot \vec \nabla \chi_{\I}(\vec x)\end{array}
\right),
\end{equation}
Self-adjointness requires equality of the domains, $D(H^\dagger) = D(H)$, which
thus implies
\begin{eqnarray}
\Gamma(\vec x)^{-1}&=&\frac{1}{\Gamma_{11}(\vec x) \Gamma_{22}(\vec x) -
\Gamma_{12}(\vec x) \Gamma_{21}(\vec x)}
\left(\begin{array}{cc} \Gamma_{22}(\vec x) & - \Gamma_{12}(\vec x) \\
- \Gamma_{21}(\vec x) & \Gamma_{11}(\vec x) \end{array}\right) \nonumber \\
&=&\left(\begin{array}{cc} \Gamma_{22}(\vec x)^* & - \Gamma_{12}(\vec x)^* \\
- \Gamma_{21}(\vec x)^* & \Gamma_{11}(\vec x)^* \end{array}\right).
\end{eqnarray}
This condition is satisfied by a 4-parameter family of self-adjoint extensions
which satisfy 
\begin{equation}
\Gamma_{ij}(\vec x) \in \exp(i \theta(\vec x)) \, \R, \quad
\Gamma_{11}(\vec x) \Gamma_{22}(\vec x) -
\Gamma_{12}(\vec x) \Gamma_{21}(\vec x) = \exp(2 i \theta(\vec x)),
\end{equation}
i.e.\ $\Gamma(\vec x)$ is real with determinant 1, up to an overall phase
$\exp(i \theta(\vec x))$. 

Using this form of $\Gamma(\vec x)$, it is straightforward to show that
self-adjointness implies probability current conservation, i.e.
\begin{eqnarray}
i \vec n(\vec x) \cdot \vec j_{\I}(\vec x)&=& 
\vec n(\vec x) \cdot \frac{1}{2 m_{\I}} \left[
\Psi_{\I}(\vec x)^* \vec \nabla \Psi_{\I}(\vec x) -
\vec \nabla \Psi_{\I}(\vec x)^* \Psi_{\I}(\vec x)\right] \nonumber \\
&=&\vec n(\vec x) \cdot \frac{1}{2 m_{\II}}
\Psi_{\II}(\vec x)^* \vec \nabla \Psi_{\II}(\vec x) 
\left[\Gamma_{11}(\vec x)^* \Gamma_{22}(\vec x) -
\Gamma_{21}(\vec x)^* \Gamma_{12}(\vec x)\right] \nonumber \\
&+&\vec n(\vec x) \cdot \frac{1}{2 m_{\II}}
\vec \nabla \Psi_{\II}(\vec x)^* \Psi_{\II}(\vec x) 
\left[\Gamma_{12}(\vec x)^* \Gamma_{21}(\vec x) -
\Gamma_{22}(\vec x)^* \Gamma_{11}(\vec x)\right] \nonumber \\
&+&\vert\Psi_{\II}(\vec x)\vert^2 
\left[\Gamma_{11}(\vec x)^* \Gamma_{21}(\vec x) -
\Gamma_{21}(\vec x)^* \Gamma_{11}(\vec x)\right] \nonumber \\
&+&\vert\vec n(\vec x) \cdot \frac{1}{2 m_{\II}} \vec \nabla \Psi_{\II}(\vec x)
\vert^2 \left[\Gamma_{12}(\vec x)^* \Gamma_{22}(\vec x) -
\Gamma_{22}(\vec x)^* \Gamma_{12}(\vec x)\right] \nonumber \\
&=&\vec n(\vec x) \cdot \frac{1}{2 m_{\II}} 
\left[\Psi_{\II}(\vec x)^* \vec \nabla \Psi_{\II}(\vec x) -
\vec \nabla \Psi_{\II}(\vec x)^* \Psi_{\II}(\vec x)\right] \nonumber \\
&=&i \vec n(\vec x) \cdot \vec j_{\II}(\vec x). 
\end{eqnarray}
Here we have used again that $\Gamma(\vec x)$ is real with determinant 1 up to
the overall phase $\exp(i \theta(\vec x))$, which implies
\begin{eqnarray}
&&\Gamma_{11}(\vec x)^* \Gamma_{22}(\vec x) -
\Gamma_{21}(\vec x)^* \Gamma_{12}(\vec x) = 1, \nonumber \\
&&\Gamma_{12}(\vec x)^* \Gamma_{21}(\vec x) -
\Gamma_{22}(\vec x)^* \Gamma_{11}(\vec x) = - 1, \nonumber \\
&&\Gamma_{11}(\vec x)^* \Gamma_{21}(\vec x) -
\Gamma_{21}(\vec x)^* \Gamma_{11}(\vec x) = 0, \nonumber \\
&&\Gamma_{12}(\vec x)^* \Gamma_{22}(\vec x) -
\Gamma_{22}(\vec x)^* \Gamma_{12}(\vec x) = 0.
\end{eqnarray}

\section{Reflecting Walls for Relativistic Fermions}

In this section we consider perfectly reflecting boundary conditions for
relativistic fermions. Such boundary conditions have been introduced in the MIT
bag model to mimic the confinement of quarks and gluons inside hadrons. Here we 
construct the most general perfectly reflecting boundary conditions for 
relativistic Dirac fermions both in one and in three spatial dimensions, as 
well as for the resulting non-relativistic fermions described by the Pauli 
equation.

\subsection{Reflecting Boundary Conditions for 1-d Dirac Fermions}

General boundary conditions for a Dirac particle in a box have been investigated
in \cite{Alo97}. Here we consider a Dirac fermion moving on the positive 
$x$-axis with a perfectly reflecting wall at $x = 0$. The corresponding free 
particle Hamiltonian is then given by
\begin{equation}
\label{alphabeta}
H = \alpha p c + \beta m c^2, \quad
\alpha = \left(\begin{array}{cc} 0 & 1 \\ 1 & 0 \end{array}\right), \quad
\beta = \left(\begin{array}{cc} 1 & 0 \\ 0 & - 1 \end{array}\right).
\end{equation}
The Hamiltonian acts on a 2-component spinor $\Psi(x,t)$, and the corresponding
continuity equation
\begin{equation}
\p_t \rho(x,t) + \p_x j(x,t) = 0,
\end{equation}
is satisfied by
\begin{equation}
\rho(x,t) = \Psi(x,t)^\dagger \Psi(x,t), \quad
j(x,t) = c \Psi(x,t)^\dagger \alpha \Psi(x,t). 
\end{equation}

Let us investigate the Hermiticity of the Hamiltonian on the positive $x$-axis
\begin{eqnarray}
\langle \chi|H|\Psi\rangle&=&\int_0^\infty dx \ \chi(x)^\dagger 
\left[- c \alpha i \p_x + \beta m c^2\right] \Psi(x) \nonumber \\
&=&\int_0^\infty dx \ \left\{\left[- c \alpha i \p_x + \beta m c^2\right]
\chi(x)\right\}^\dagger \Psi(x) - i c \chi(0)^\dagger \alpha \Psi(0) \nonumber \\
&=&\langle \Psi|H|\chi\rangle^* - i c \chi(0)^\dagger \alpha \Psi(0),
\end{eqnarray}
which thus leads to the Hermiticity condition
\begin{equation}
\label{Dirachermiticity1d}
\chi(0)^\dagger \alpha \Psi(0) = 0.
\end{equation}
We now introduce the self-adjoint extension condition
\begin{equation}
\label{Diracselfadjointness1d}
\Psi_2(0) = \lambda \Psi_1(0), \quad \lambda \in \C,
\end{equation}
which reduces eq.(\ref{Dirachermiticity1d}) to
\begin{equation}
\chi(0)^\dagger 
\left(\begin{array}{cc} 0 & 1 \\ 1 & 0 \end{array}\right) \Psi(0) =
\left[\chi_1(0)^* \lambda + \chi_2(0)^*\right] \Psi_1(0) = 0 \ 
\Rightarrow \ \chi_2(0) = - \lambda^* \chi_1(0).
\end{equation}
In order to guarantee self-adjointness of $H$, i.e.\ $D(H) = D(H^\dagger)$, we 
must request
\begin{equation}
\lambda = - \lambda^*,
\end{equation}
i.e.\ $\lambda$ must be purely imaginary. For 1-d Dirac fermions, there is thus 
a 1-parameter family of self-adjoint extensions that characterizes a perfectly 
reflecting wall. The self-adjointness condition 
eq.(\ref{Diracselfadjointness1d}) implies
\begin{eqnarray}
j(0)&=&c \Psi(0)^\dagger \alpha \Psi(0) = c \Psi(0)^\dagger 
\left(\begin{array}{cc} 0 & 1 \\ 1 & 0 \end{array}\right) \Psi(0) = 
c \left[\Psi_1(0)^* \Psi_2(0) + \Psi_2(0)^* \Psi_1(0)\right] \nonumber \\
&=&c \left[\Psi_1(0)^* \lambda \Psi_1(0) + \Psi_1(0)^* \lambda^* \Psi_1(0)\right]
= 0.
\end{eqnarray}

In the chiral limit, $m = 0$, not only the vector current but also the axial 
current is conserved. In the basis we have chosen, the $\gamma$-matrices take 
the form
\begin{equation}
\label{gamma2d}
\gamma^0 = \beta = \left(\begin{array}{cc} 1 & 0 \\ 0 & - 1 \end{array}\right),
\quad
\gamma^1 = \gamma^0 \alpha = 
\left(\begin{array}{cc} 0 & 1 \\ - 1 & 0 \end{array}\right), \quad
\gamma^3 = \gamma^0 \gamma^1 = 
\left(\begin{array}{cc} 0 & 1 \\ 1 & 0 \end{array}\right).
\end{equation}
Thus the axial charge density and the axial current density are given by
\begin{eqnarray}
&&\rho_A(x,t) = \Psi(x,t)^\dagger \gamma^3 \Psi(x,t), \nonumber \\
&&j_A(x,t) = c \Psi(x,t)^\dagger \gamma^0 \gamma^1 \gamma^3 \Psi(x,t) = 
c \Psi(x,t)^\dagger \Psi(x,t).
\end{eqnarray}
Indeed, it is easy to convince oneself that 
$\p_t \rho_A(x,t) + \p_x j_A(x,t) = 0$ in the chiral limit $m = 0$. Let us 
now consider the axial current at the boundary $x = 0$
\begin{equation}
j_A(0) = c \left[|\Psi_1(0)|^2 + |\Psi_2(0)|^2\right] = 
c \left(1 + |\lambda|^2\right) |\Psi_1(0)|^2 \geq 0.
\end{equation}
Since in general $\Psi_1(0) \neq 0$, the axial current does not vanish at the
boundary. Hence, chiral symmetry is explicitly broken by the most general
perfectly reflecting boundary condition. It is well-known that this is indeed 
the case for the boundary condition in the MIT bag model 
\cite{Cho74,Cho74a,Has78}. Only in the chiral bag model the axial current is
conserved because it is carried by a pion field outside the bag 
\cite{Ino75,Bro79}.

It is interesting to ask how the self-adjoint extension parameter $\lambda$ in
the relativistic case is related to the parameter $\gamma$ in the 
non-relativistic limit, in which
\begin{equation}
\Psi_2(x) = \frac{p}{2 m c} \Psi_1(x) = \frac{1}{2 m c i} \p_x \Psi_1(x).
\end{equation}
Inserting this relation in the relativistic current,
\begin{eqnarray}
j(x)&=&c \Psi(x)^\dagger \alpha \Psi(x) = 
c \left[\Psi_1(x)^* \Psi_2(x) + \Psi_2(x)^* \Psi_1(x)\right] 
\nonumber \\
&=&\frac{1}{2 m i} \left[\Psi_1(x)^* \p_x \Psi_1(x) - \p_x \Psi_1(x)^*
\Psi_1(x)\right],
\end{eqnarray}
we indeed recover the non-relativistic current of eq.(\ref{current}). In
the non-relativistic limit, the relativistic self-adjointness condition 
eq.(\ref{Diracselfadjointness1d}) takes the form
\begin{equation}
\frac{1}{2 m c \, i} \p_x \Psi_1(0) = \lambda \Psi_1(0) \ \Rightarrow \
- 2 m c \, i \lambda \Psi_1(0) +  \p_x \Psi_1(0) = 0.
\end{equation}
Hence, the non-relativistic self-adjoint extension parameter of 
eq.(\ref{wallboundary}) can be identified as
\begin{equation}
\gamma = - 2 m c \, i \lambda,
\end{equation}
which is indeed real because $\lambda$ is purely imaginary.

\subsection{Reflecting Boundary Conditions for 3-d Dirac Fermions}

Let us now consider Dirac fermions coupled to an external static 
electromagnetic field and confined to a finite domain $\Omega$. The 
corresponding Hamiltonian then takes the form
\begin{eqnarray}
&&H = \vec \alpha \cdot \left(\vec p c + e \vec A(\vec x)\right) + \beta m c^2 -
e \Phi(\vec x) = - i \vec \alpha \cdot \vec D c + \beta m c^2 - e \Phi(\vec x),
\nonumber \\
&&\vec \alpha = \left(\begin{array}{cc} \0 & \vec \sigma \\ \vec \sigma & \0
\end{array}\right), \quad
\beta = \left(\begin{array}{cc} \1 & \0 \\ \0 & - \1 \end{array}\right).
\end{eqnarray}
Here $\1$ and $\0$ are the $2 \times 2$ unit- and zero-matrix, and
$\vec \sigma$ is the vector of Pauli matrices, while $\Phi(\vec x)$ and
$\vec A(\vec x)$ are the scalar and vector potential, and $e$ is the electric
charge. The covariant derivative is given by
\begin{equation}
\vec D = \vec \nabla + i \frac{e}{c} \vec A(\vec x).
\end{equation}
The Dirac Hamiltonian acts on a 4-component spinor $\Psi(\vec x,t)$. In 
this case, the continuity equation
\begin{equation}
\p_t \rho(\vec x,t) + \vec \nabla \cdot \vec j(\vec x,t) = 0,
\end{equation}
is satisfied by
\begin{equation}
\rho(\vec x,t) = \Psi(\vec x,t)^\dagger \Psi(\vec x,t), \quad
\vec j(\vec x,t) = c \Psi(\vec x,t)^\dagger \vec \alpha \Psi(\vec x,t). 
\end{equation}
Under time-independent gauge transformations, the gauge fields as well as the 
Dirac spinor transform as
\begin{equation}
^\varphi\Phi(\vec x) = \Phi(\vec x), \ 
^\varphi\vec A(\vec x) = \vec A(\vec x) - \vec \nabla \varphi(\vec x), \
^\varphi\Psi(\vec x) = \exp\left(i \frac{e}{c} \varphi(\vec x)\right) 
\Psi(\vec x).
\end{equation}

In order to investigate the Hermiticity of the Hamiltonian in the finite 
spatial domain $\Omega$, we consider
\begin{eqnarray}
\langle \chi|H|\Psi\rangle&=&\int_\Omega d^3x \ \chi(\vec x)^\dagger 
\left[\vec \alpha \cdot \left(- i c \vec \nabla + e \vec A(\vec x)\right) +
\beta m c^2 - e \Phi(\vec x)\right] \Psi(\vec x) \nonumber \\
&=&\int_\Omega d^3x \ \left\{\left[\vec \alpha \cdot 
\left(- i c \vec \nabla + e \vec A(\vec x)\right) + \beta m c^2 - e \Phi(\vec x)
\right]\chi(\vec x)\right\}^\dagger \Psi(\vec x) \nonumber \\
&-&i c \int_{\p \Omega} d\vec n \cdot \chi(\vec x)^\dagger \vec \alpha 
\Psi(\vec x) \nonumber \\
&=&\langle \Psi|H|\chi\rangle^* - 
i c \int_{\p \Omega} d\vec n \cdot \chi(\vec x)^\dagger \vec \alpha \Psi(\vec x),
\end{eqnarray}
which thus leads to the Hermiticity condition
\begin{equation}
\label{Dirachermiticity}
\chi(\vec x)^\dagger \vec n(\vec x) \cdot \vec \alpha \Psi(\vec x) = 0, 
\quad \vec x \in \p \Omega.
\end{equation}
We now introduce the self-adjoint extension condition
\begin{equation}
\label{Diracselfadjointness}
\left(\begin{array}{c} \Psi_3(\vec x) \\ \Psi_4(\vec x) \end{array}\right) = 
\lambda(\vec x) 
\left(\begin{array}{c} \Psi_1(\vec x) \\ \Psi_2(\vec x) \end{array}\right),
\quad \lambda(\vec x) \in GL(2,\C), \quad \vec x \in \p \Omega,
\end{equation}
which reduces eq.(\ref{Dirachermiticity}) to
\begin{eqnarray}
&&\chi(\vec x)^\dagger 
\left(\begin{array}{cc} \0 & \vec n(\vec x) \cdot \vec \sigma \\ 
\vec n(\vec x) \cdot \vec \sigma & \0 \end{array}\right) \Psi(\vec x) = 
\nonumber \\
&&\left[\left(\chi_1(\vec x)^*,\chi_2(\vec x)^* \right) 
\vec n(\vec x) \cdot \vec \sigma \lambda(\vec x) + 
\left(\chi_3(\vec x)^*,\chi_4(\vec x)^* \right)
\vec n(\vec x) \cdot \vec \sigma \right]
\left(\begin{array}{c} \Psi_1(\vec x) \\ \Psi_2(\vec x) \end{array}\right) 
= 0 \ \Rightarrow \nonumber \\ 
&&\left(\begin{array}{c} \chi_3(\vec x) \\ \chi_4(\vec x) \end{array}\right) = 
- \vec n(\vec x) \cdot \vec \sigma \lambda(\vec x)^\dagger 
\vec n(\vec x) \cdot \vec \sigma
\left(\begin{array}{c} \chi_1(\vec x) \\ \chi_2(\vec x) \end{array}\right),
\end{eqnarray}
In order to guarantee self-adjointness of $H$, i.e.\ $D(H) = D(H^\dagger)$, we 
now demand
\begin{equation}
\lambda(\vec x) = - \vec n(\vec x) \cdot \vec \sigma \lambda(\vec x)^\dagger 
\vec n(\vec x) \cdot \vec \sigma \ \Rightarrow \
\vec n(\vec x) \cdot \vec \sigma \lambda(\vec x) = -
\left[\vec n(\vec x) \cdot \vec \sigma \lambda(\vec x)\right]^\dagger.
\end{equation}
Hence, $\vec n(\vec x) \cdot \vec \sigma \lambda(\vec x)$ is anti-Hermitean. 
For Dirac fermions, there is thus a 4-parameter family of self-adjoint 
extensions that characterizes a perfectly reflecting wall. It is important to
note that the self-adjointness condition eq.(\ref{Diracselfadjointness}) is 
gauge covariant and implies
\begin{eqnarray}
\vec n(\vec x) \cdot \vec j(\vec x)\!\!\!\!&=&\!\!\!\!c \Psi(\vec x)^\dagger 
\left(\begin{array}{cc} \0 & \vec n(\vec x) \cdot \vec \sigma \\ 
\vec n(\vec x) \cdot \vec \sigma & \0 \end{array}\right) \Psi(\vec x) = 
\nonumber \\
&=&\!\!\!\!c \left[\left(\Psi_1(\vec x)^*,\Psi_2(\vec x)^* \right) 
\vec n(\vec x) \cdot \vec \sigma \lambda(\vec x) + 
\left(\Psi_3(\vec x)^*,\Psi_4(\vec x)^* \right)
\vec n(\vec x) \cdot \vec \sigma \right]
\left(\begin{array}{c} \Psi_1(\vec x) \\ \Psi_2(\vec x) \end{array}\right) 
\nonumber \\
&=&\!\!\!\!c \left(\Psi_1(\vec x)^*,\Psi_2(\vec x)^* \right) \left[
\vec n(\vec x) \cdot \vec \sigma \lambda(\vec x) + 
\lambda(\vec x)^\dagger \vec n(\vec x) \cdot \vec \sigma \right]
\left(\begin{array}{c} \Psi_1(\vec x) \\ \Psi_2(\vec x) \end{array}\right) 
= 0.
\end{eqnarray}

Let us again consider the chiral limit $m = 0$, in which the axial current
is also conserved. The $\gamma$-matrices now take the form
\begin{equation}
\label{gamma4d}
\gamma^0 = \beta = 
\left(\begin{array}{cc} \1 & \0 \\ \0 & -\1 \end{array}\right),
\quad
\vec \gamma = \gamma^0 \vec \alpha = 
\left(\begin{array}{cc} \0 & \vec \sigma \\ - \vec \sigma & \0 \end{array}
\right), \quad
\gamma^5 = i \gamma^0 \gamma^1 \gamma^2 \gamma^3 = 
\left(\begin{array}{cc} \0 & \1 \\ \1 & \0 \end{array}\right).
\end{equation}
Hence the axial charge density and the axial current density are given by
\begin{eqnarray}
&&\rho_A(\vec x,t) = \Psi(\vec x,t)^\dagger \gamma^5 \Psi(\vec x,t), 
\nonumber \\
&&\vec j_A(\vec x,t) =
c \Psi(\vec x,t)^\dagger \gamma^0 \vec \gamma \gamma^5 \Psi(\vec x,t) = 
- c \Psi(\vec x,t)^\dagger 
\left(\begin{array}{cc} \vec \sigma & \0 \\ \0 & \vec \sigma \end{array}\right) 
\Psi(\vec x,t), 
\end{eqnarray}
such that the axial current at the boundary is
\begin{eqnarray}
\vec n(\vec x) \cdot \vec j_A(\vec x)\!\!\!\!&=&\!\!\!\!- c \Psi(\vec x)^\dagger 
\left(\begin{array}{cc} \vec n(\vec x) \cdot \vec \sigma & \0 \\ 
\0 & \vec n(\vec x) \cdot \vec \sigma \end{array}\right) \Psi(\vec x) = 
\nonumber \\
&=&\!\!\!\!- c \left(\Psi_1(\vec x)^*,\Psi_2(\vec x)^* \right) 
\left[\vec n(\vec x) \cdot \vec \sigma + 
\lambda(\vec x)^\dagger \vec n(\vec x) \cdot \vec \sigma \lambda(\vec x)\right]
\left(\begin{array}{c} \Psi_1(\vec x) \\ \Psi_2(\vec x) \end{array}\right).
\quad \quad
\end{eqnarray}
As in the 1-d case, in general, the axial current does not vanish at the 
boundary. Hence, chiral symmetry is again explicitly broken by the most general 
perfectly reflecting boundary condition.

\subsection{Reflecting Boundary Conditions in the Non-relativistic Limit}

In the non-relativistic limit, the lower components of the Dirac spinor are 
given by
\begin{equation}
\label{nrlimit}
\left(\begin{array}{c} \Psi_3(\vec x) \\ \Psi_4(\vec x) \end{array} \right) =
\frac{\vec \sigma \cdot \left(\vec p c + e \vec A(\vec x)\right)}{2 m c^2} 
\left(\begin{array}{c} \Psi_1(\vec x) \\ \Psi_2(\vec x) \end{array} \right) =
\frac{1}{2 m c \, i} \vec \sigma \cdot \vec D \Psi(\vec x), 
\end{equation}
with the 2-component Pauli spinor
\begin{equation}
\Psi(\vec x) =
\left(\begin{array}{c} \Psi_1(\vec x) \\ \Psi_2(\vec x) \end{array} \right).
\end{equation}
The Dirac equation then reduces to the Pauli equation. Expanding up to the 
leading Zeemann term, but neglecting the higher order spin-orbit coupling and 
Darwin terms, the Hamiltonian entering the Pauli equation takes the form
\begin{equation}
\label{PauliH}
H = m c^2 + \frac{\left(\vec p c + e \vec A(\vec x)\right)^2}{2 m c^2} - 
e \Phi(\vec x) + \mu \vec \sigma \cdot \vec B(\vec x),
\end{equation}
where $\vec B(\vec x) = \vec \nabla \times \vec A(\vec x)$ is the magnetic 
field and $\mu = e/2 m c$ is the magnetic moment of the fermion.

Using eq.(\ref{nrlimit}), the relativistic current reduces to
\begin{eqnarray}
\vec j(\vec x)&=&c 
\left(\Psi_1(\vec x)^*,\Psi_2(\vec x)^*,\Psi_3(\vec x)^*,\Psi_4(\vec x)^*\right)
\left(\begin{array}{cc} \0 & \vec \sigma \\ \vec \sigma & \0 \end{array}\right)
\left(\begin{array}{c} \Psi_1(\vec x) \\ \Psi_2(\vec x) \\ 
\Psi_3(\vec x) \\ \Psi_4(\vec x) \end{array} \right) \nonumber \\
&=&\frac{1}{2 m i} \left[\Psi(\vec x)^\dagger \vec D \Psi(\vec x) - 
(\vec D \Psi(\vec x))^\dagger \Psi(\vec x)\right] \nonumber \\
&-&\frac{1}{2 m} 
\left[\Psi(\vec x,t)^\dagger \vec \sigma \times \vec D \Psi(\vec x,t) - 
(\vec D \Psi(\vec x,t))^\dagger \times \vec \sigma \Psi(\vec x,t)\right]
\nonumber \\
&=&\frac{1}{2 m i} 
\left[\Psi(\vec x,t)^\dagger \vec D \Psi(\vec x,t) - 
(\vec D \Psi(\vec x,t))^\dagger \Psi(\vec x,t)\right] \nonumber \\
&+&\frac{1}{2 m} \vec \nabla \times 
\left[\Psi(\vec x,t)^\dagger \vec \sigma \Psi(\vec x,t)\right].
\end{eqnarray}
As a curl, the spin term entering the current is automatically divergenceless. 
In this case, the continuity equation 
\begin{equation}
\p_t \rho(\vec x,t) + \vec \nabla \cdot \vec j(\vec x,t) = 0,
\end{equation}
is satisfied with the probability density 
$\rho(\vec x,t) = \Psi(\vec x,t)^\dagger \Psi(\vec x,t)$.

Introducing the gauge covariant boundary condition
\begin{equation}
\label{sadjnonrel}
\gamma(\vec x) \Psi(\vec x) + \vec n(\vec x) \cdot 
\left[\vec D  \Psi(\vec x) - i \vec \sigma \times \vec D  \Psi(\vec x)\right] = 
0, \quad \gamma(\vec x) \in GL(2,\C), \quad \vec x \in \p \Omega,
\end{equation}
we obtain
\begin{eqnarray}
\vec n(\vec x) \cdot \vec j(\vec x)&=&\frac{1}{2 m i} 
\left[\Psi(\vec x)^\dagger \vec n(\vec x) \cdot \vec D \Psi(\vec x) - 
(\vec n(\vec x) \cdot \vec D \Psi(\vec x))^\dagger \Psi(\vec x)\right] 
\nonumber \\
&-&\frac{1}{2 m} \left[\Psi(\vec x,t)^\dagger \vec n \cdot 
\left(\vec \sigma \times \vec D \Psi(\vec x,t)\right) - 
\vec n \cdot \left((\vec D \Psi(\vec x,t))^\dagger \times \vec \sigma\right) 
\Psi(\vec x,t)\right] \nonumber \\
&=&\frac{1}{2 m i} 
\left[- \Psi(\vec x)^\dagger \gamma(\vec x) \Psi(\vec x) + 
\Psi(\vec x)^\dagger \gamma(\vec x)^\dagger \Psi(\vec x)\right] = 0,
\end{eqnarray}
which immediately implies
\begin{equation}
\gamma(\vec x)^\dagger = \gamma(\vec x).
\end{equation}
Hence, again there is a 4-parameter family of self-adjoint extensions, now 
parameterized by a $2 \times 2$ Hermitean matrix.

In complete analogy to the previous cases, by partial integration one arrives
at the Hermiticity condition for the Pauli Hamiltonian of eq.(\ref{PauliH})
\begin{equation}
\label{Hnonrel}
\int_{\p \Omega}  d\vec n \cdot \left[\left(\vec D \chi(\vec x)\right)^\dagger 
\Psi(\vec x) - \chi(\vec x)^\dagger \vec D \Psi(\vec x)\right] = 0.
\end{equation}
One also readily derives
\begin{equation}
\left(\vec D \chi(\vec x)\right)^\dagger \times \vec \sigma \Psi(\vec x) -
\chi(\vec x)^\dagger \vec \sigma \times \vec D \Psi(\vec x) =
\vec \nabla \times \left(\chi(\vec x)^\dagger \vec \sigma \Psi(\vec x)\right).
\end{equation}
Using Stoke's theorem as well as $\p(\p \Omega) = \emptyset$, (i.e.\ the 
boundary of a boundary is an empty set), one then obtains
\begin{eqnarray}
&&\int_{\p \Omega}  d\vec n \cdot 
\left[\left(\vec D \chi(\vec x)\right)^\dagger \times \vec \sigma \Psi(\vec x) -
\chi(\vec x)^\dagger \vec \sigma \times \vec D \Psi(\vec x)\right] = \nonumber \\
&&\int_{\p \Omega}  d\vec n \cdot 
\vec \nabla \times \left(\chi(\vec x)^\dagger \vec \sigma \Psi(\vec x)\right) =
\int_{\p(\p \Omega)} d\vec l \cdot \chi(\vec x)^\dagger \vec \sigma \Psi(\vec x) = 
0,
\end{eqnarray}
such that the Hermiticity condition eq.(\ref{Hnonrel}) may be rewritten as
\begin{equation}
\int_{\p \Omega}  d\vec n \cdot \left[\left(\vec D \chi(\vec x) - 
i \vec \sigma \times \vec D \chi(\vec x)\right)^\dagger \Psi(\vec x) -
\chi(\vec x)^\dagger \left(\vec D \Psi(\vec x) - i \vec \sigma \times 
\vec D \Psi(\vec x) \right)\right] = 0.
\end{equation}
Using the self-adjointness condition eq.(\ref{sadjnonrel}), this relation
reduces to
\begin{equation}
\int_{\p \Omega}  d^2x \ \left[\chi(\vec x)^\dagger \gamma(\vec x)^\dagger 
\Psi(\vec x) - \chi(\vec x)^\dagger \gamma(\vec x) \Psi(\vec x) \right] = 0,
\end{equation}
which is indeed satisfied because $\gamma(\vec x)$ is Hermitean.

It is again interesting to ask how the Hermitean matrix $\gamma(\vec x)$ 
emerges from the matrix $\lambda(\vec x)$ in the non-relativistic limit.
Noting that the self-adjointness condition eq.(\ref{sadjnonrel}) can also be
expressed as
\begin{equation}
\gamma(\vec x) \Psi(\vec x) + \vec n(\vec x) \cdot \vec \sigma \,
\vec \sigma \cdot \vec D \Psi(\vec x) =
\gamma(\vec x) \Psi(\vec x) + 2 m c \, i \vec n(\vec x) \cdot \vec \sigma 
\lambda(\vec x) \Psi(\vec x) = 0,
\end{equation}
one immediately identifies
\begin{equation}
\gamma(\vec x) = - 2 m c \, i \, \vec n(\vec x) \cdot \vec \sigma 
\lambda(\vec x).
\end{equation}
Since $\vec n(\vec x) \cdot \vec \sigma \lambda(\vec x)$ is anti-Hermitean, the 
resulting matrix $\gamma(\vec x)$ is indeed Hermitean.

\section{Domain Wall Fermions}

Using Shamir's variant \cite{Sha93} of Kaplan's domain wall fermions 
\cite{Kap92}, let us imagine that our world has an additional hidden spatial 
dimension and that we live very near a perfectly reflecting flat domain wall. 
For simplicity, we first explore this idea in $(2+1)$-d and then extend it
to $(4+1)$-d. Since there is no notion of chirality in odd space-time
dimensions, it is natural to consider massive Dirac fermions.

\subsection{Domain Wall Boundary Conditions in $(2+1)$-d}

Let us consider fermions moving freely along the $x_1$-direction and localized 
near a perfectly reflecting wall located at $x_3 = 0$ at very small positive 
values of $x_3$. We denote the two spatial coordinates by $x_1$ and $x_3$, 
while time is denoted by $x_0$. We would reserve $x_2$ for Euclidean time 
which, however, does not play a role in the present paper. Starting from
the $\gamma$-matrices of eq.(\ref{gamma2d}), it is more convenient to change to 
a chiral basis by performing the unitary transformation
\begin{eqnarray}
&&\!\!\!\!\!\!\!\!\!\!\widetilde \gamma^0 = U \gamma^0 U^\dagger = 
\left(\begin{array}{cc} 0 & 1 \\ 1 & 0 \end{array}\right), \quad 
\widetilde \gamma^1 =  U \gamma^1 U^\dagger = 
\left(\begin{array}{cc} 0 & -1 \\ 1 & 0 \end{array}\right), \nonumber \\
&&\!\!\!\!\!\!\!\!\!\!\widetilde \gamma^3 =  U \gamma^3 U^\dagger = 
\left(\begin{array}{cc} 1 & 0 \\ 0 & -1 \end{array}\right), \quad
U = \frac{1}{\sqrt{2}} 
\left(\begin{array}{cc} 1 & 1 \\ 1 & -1 \end{array}\right) = U^\dagger,
\nonumber \\
&&\!\!\!\!\!\!\!\!\!\!\widetilde \alpha = 
\widetilde \gamma^0 \widetilde \gamma^1 = 
\left(\begin{array}{cc} 1 & 0 \\ 0 & -1 \end{array}\right), \quad 
\widetilde \beta =  \widetilde \gamma^0 = 
\left(\begin{array}{cc} 0 & 1 \\ 1 & 0 \end{array}\right), \quad
\widetilde \alpha^3 = - i \widetilde \gamma^0 \widetilde \gamma^3 = 
i \left(\begin{array}{cc} 0 & 1 \\ -1 & 0 \end{array}\right). \nonumber \\ \,
\end{eqnarray}
The domain wall fermion Hamiltonian then takes the form
\begin{equation}
H = \widetilde \alpha p c + \widetilde \beta m c^2 + \widetilde \alpha^3 p_3 c =
\widetilde \alpha p c + \widetilde \beta m c^2 - i \widetilde \alpha^3 c \p_3,
\end{equation}
and the 3-component of the current is given by
\begin{eqnarray}
j^3(x_1,x_3)&=&c \Psi(x_1,x_3)^\dagger \widetilde \alpha^3 \Psi(x_1,x_3) 
\nonumber \\
&=&i c \left(\Psi_R(x_1,x_3)^*,\Psi_L(x_1,x_3)^*\right)
\left(\begin{array}{cc} 0 & 1 \\ -1 & 0 \end{array}\right)
\left(\begin{array}{c} \Psi_R(x_1,x_3) \\ \Psi_L(x_1,x_3) \end{array}\right).
\end{eqnarray}
In complete analogy to the cases discussed before, the Hermiticity condition
then reads
\begin{equation}
\left(\chi_R(x_1,0)^*,\chi_L(x_1,0)^*\right)
\left(\begin{array}{cc} 0 & 1 \\ -1 & 0 \end{array}\right)
\left(\begin{array}{c} \Psi_R(x_1,0) \\ \Psi_L(x_1,0) \end{array}\right) = 0.
\end{equation}
Introducing the self-adjointness condition
\begin{equation}
\label{boundwall}
\Psi_R(x_1,0) = \eta \Psi_L(x_1,0),
\end{equation}
the Hermiticity condition turns into
\begin{equation}
\left(\chi_R(x_1,0)^* - \chi_L (x_1,0)^* \eta\right) \Psi_L(x_1,0) = 0 \
\Rightarrow \ \chi_R(x_1,0) = \eta^* \chi_L(x_1,0).
\end{equation}
Thus, in order to ensure $D(H^\dagger) = D(H)$ and hence self-adjointness, we
must demand
\begin{equation}
\eta = \eta^* \in \R.
\end{equation}
Since the boundary condition eq.(\ref{boundwall}) couples left- and right-handed
components, which transform differently under the $(1+1)$-d Lorentz group, it
explicitly breaks $(1+1)$-d Lorentz invariance (unless $\eta = 0$ or 
$\pm \infty$). 

The Dirac equation takes the form
\begin{equation}
\left(\begin{array}{cc} p c & m c^2 + c \p_3 \\ m c^2 - c \p_3 & - p c 
\end{array}\right) 
\left(\begin{array}{c} \Psi_R(x_1,x_3) \\ \Psi_L(x_1,x_3) \end{array}\right) =
E \left(\begin{array}{c} \Psi_R(x_1,x_3) \\ \Psi_L(x_1,x_3) \end{array}\right).
\end{equation}
Inserting the ansatz
\begin{equation}
\Psi_R(x_1,x_3) = A_R \exp(i p x_1) \exp(- \varkappa x_3), \quad
\Psi_L(x_1,x_3) = A_L \exp(i p x_1) \exp(- \varkappa x_3),
\end{equation}
for a state localized on the domain wall, the Dirac equation reduces to
\begin{equation}
\left(\begin{array}{cc} p c & m c^2 - c \varkappa \\ m c^2 + c \varkappa & 
- p c \end{array}\right) \left(\begin{array}{c} A_R \\ A_L \end{array}\right) =
E \left(\begin{array}{c} A_R \\ A_L \end{array}\right).
\end{equation}
Imposing the boundary condition eq.(\ref{boundwall}), one then obtains
\begin{equation}
E = \frac{2 \eta}{1 + \eta^2} m c^2 - 
\frac{1 - \eta^2}{1 + \eta^2} p c, \quad
c \varkappa = \frac{1 - \eta^2}{1 + \eta^2} m c^2 + 
\frac{2 \eta}{1 + \eta^2} p c.
\end{equation}
This is the energy-momentum dispersion relation of a fermion moving with the 
speed
\begin{equation}
v = \left\vert\frac{1 - \eta^2}{1 + \eta^2}\right\vert c \leq c,
\end{equation}
and coupled to a chemical potential
\begin{equation}
\mu = \frac{2 \eta}{1 + \eta^2} m c^2.
\end{equation}
Only for $\eta = 0$ or $\pm \infty$, one obtains $v = c$ and $\mu = 0$, as
a consequence of $(1+1)$-d Lorentz invariance. It is important to note that
normalizability of the state localized on the domain wall requires 
$\varkappa > 0$, which restricts the allowed range of $p$. Only for $\eta = 0$
or $\pm \infty$, there is no restriction and one obtains a relativistic massless
left-handed domain wall fermion with $E = - p c$. This means that particles 
(i.e.\ positive energy states) are left-moving ($p < 0$) while anti-particles 
are right-moving.

\subsection{Domain Wall Boundary Condition in $(4+1)$-d}

Let us now consider domain wall fermions in $(4+1)$-d localized near a 
perfectly reflecting wall located at $x_5 = 0$ at very small positive 
values of $x_5$. We denote the four spatial coordinates by $x_1, x_2, x_3$ and 
$x_5$, reserving $x_4$ for Euclidean time. In this case, the 
$\gamma$-matrices are given by eq.(\ref{gamma4d}). Again, it is useful to 
change to a chiral basis
\begin{eqnarray}
&&\!\!\!\!\!\!\!\!\!\!\widetilde \gamma^0 = U \gamma^0 U^\dagger = 
\left(\begin{array}{cc} \0 & \1 \\ \1 & \0 \end{array}\right), \quad 
\widetilde{\vec \gamma} =  U \vec \gamma U^\dagger = 
\left(\begin{array}{cc} \0 & - \vec \sigma \\ \vec \sigma & \0 
\end{array}\right), \nonumber \\ 
&&\!\!\!\!\!\!\!\!\!\!\widetilde \gamma^5 =  U \gamma^5 U^\dagger = 
\left(\begin{array}{cc} \1 & \0 \\ \0 & - \1 \end{array}\right),  \quad
U = \frac{1}{\sqrt{2}} 
\left(\begin{array}{cc} \1 & \1 \\ \1 & - \1 \end{array}\right) = U^\dagger,
\nonumber \\
&&\!\!\!\!\!\!\!\!\!\!\widetilde{\vec \alpha} = 
\widetilde \gamma^0 \widetilde{\vec \gamma} = 
\left(\begin{array}{cc} \vec \sigma & \0 \\ \0 & - \vec \sigma 
\end{array}\right), \quad 
\widetilde \beta =  \widetilde \gamma^0 = 
\left(\begin{array}{cc} \0 & \1 \\ \1 & \0 \end{array}\right), \quad
\widetilde \alpha^5 = - i \widetilde \gamma^0 \widetilde \gamma^5 = 
i \left(\begin{array}{cc} \0 & \1 \\ - \1 & \0 \end{array}\right). 
\nonumber \\ \,
\end{eqnarray}
The domain wall fermion Hamiltonian is now given by
\begin{equation}
H = \widetilde{\vec \alpha} \cdot \vec p c + \widetilde \beta m c^2 + 
\widetilde \alpha^5 p_5 c =
\widetilde{\vec \alpha} \cdot \vec p c + \widetilde \beta m c^2 - i 
\widetilde \alpha^5 c \p_5,
\end{equation}
and the 5-component of the current takes the form
\begin{eqnarray}
j^5(\vec x,x_5)&=&c \Psi(\vec x,x_5)^\dagger \widetilde \alpha^5 \Psi(\vec x,x_5) 
\nonumber \\
&=&i c \left(\Psi_R(\vec x,x_5)^\dagger,\Psi_L(\vec x,x_5)^\dagger\right)
\left(\begin{array}{cc} \0 & \1 \\ - \1 & \0 \end{array}\right)
\left(\begin{array}{c} \Psi_R(\vec x,x_5) \\ \Psi_L(\vec x,x_5)
\end{array}\right),
\end{eqnarray}
where $\Psi_R(\vec x,x_5)$ and $\Psi_L(\vec x,x_5)$ are 2-component Weyl 
spinors. The Hermiticity condition now reads
\begin{equation}
\label{Hermitewall}
\left(\chi_R(\vec x,0)^\dagger,\chi_L(\vec x,0)^\dagger\right)
\left(\begin{array}{cc} \0 & \1 \\ - \1 & \0 \end{array}\right)
\left(\begin{array}{c} \Psi_R(\vec x,0) \\ \Psi_L(\vec x,0) 
\end{array}\right) = 0,
\end{equation}
and the self-adjointness condition takes the form
\begin{equation}
\label{boundwall5}
\Psi_R(\vec x,0) = \eta \Psi_L(\vec x,0), \quad \eta \in GL(2,\C).
\end{equation}
Inserting this in the Hermiticity condition eq.(\ref{Hermitewall}), one obtains
\begin{equation}
\left(\chi_R(\vec x,0)^\dagger - \chi_L (\vec x,0)^\dagger \eta\right) 
\Psi_L(\vec x,0) = 0 \ \Rightarrow \ \chi_R(\vec x,0) = \eta^\dagger 
\chi_L(\vec x,0).
\end{equation}
In order to ensure $D(H^\dagger) = D(H)$, we must hence demand
\begin{equation}
\eta = \eta^\dagger = \eta_0 \1 + \vec \eta \cdot \vec \sigma.
\end{equation}
In this case, there is a 4-parameter family of self-adjoint extensions. For
general $\eta$, the boundary condition explicitly breaks $(3+1)$-d Lorentz
invariance, and for $\vec \eta \neq 0$ it even breaks 3-d spatial rotation 
invariance.

The Dirac equation now takes the form
\begin{equation}
\left(\begin{array}{cc} \vec \sigma \cdot \vec p c & m c^2 + c \p_5 \\ 
m c^2 - c \p_5 & - \vec \sigma \cdot \vec p c \end{array}\right) 
\left(\begin{array}{c} \Psi_R(\vec x,x_5) \\ \Psi_L(\vec x,x_5) 
\end{array}\right) =
E \left(\begin{array}{c} \Psi_R(\vec x,x_5) \\ \Psi_L(\vec x,x_5) 
\end{array}\right).
\end{equation}
In analogy to the $(2+1)$-d case, we make the ansatz
\begin{equation}
\Psi_R(\vec x,x_5) = A_R \exp(i \vec p \cdot \vec x) \exp(- \varkappa x_5), 
\quad
\Psi_L(\vec x,x_5) = A_L \exp(i \vec p \cdot \vec x) \exp(- \varkappa x_5),
\end{equation}
which reduces the Dirac equation to
\begin{equation}
\left(\begin{array}{cc} \sigma_3 |\vec p \,| c & m c^2 - c \varkappa \\ 
m c^2 + c \varkappa & - \sigma_3 |\vec p \,| c \end{array}\right) 
\left(\begin{array}{c} A'_R \\ A'_L \end{array}\right) =
E \left(\begin{array}{c} A'_R \\ A'_L \end{array}\right).
\end{equation}
Here we have performed a unitary transformation $U(\vec p \,)$ to diagonalize
$\vec \sigma \cdot \vec p$, i.e.
\begin{equation}
U(\vec p \,) \vec \sigma \cdot \vec p \, U(\vec p \,)^\dagger = 
|\vec p \,| \sigma_3, 
\quad A'_R = U(\vec p \,) A_R, \quad A'_L = U(\vec p \,) A_L.
\end{equation}
For simplicity, we now restrict ourselves to the rotation invariant case 
$\vec \eta = 0$, such that the boundary condition reduces to 
$A'_R = \eta_0 A'_L$ with $\eta_0 \in \R$. One then obtains
\begin{equation}
E = \frac{2 \eta_0}{1 + \eta_0^2} m c^2 \mp
\frac{1 - \eta_0^2}{1 + \eta_0^2} |\vec p \,| c, \quad
c \varkappa = \frac{1 - \eta_0^2}{1 + \eta_0^2} m c^2 \pm 
\frac{2 \eta_0}{1 + \eta_0^2} |\vec p \,| c.
\end{equation}
Again, this is the dispersion relation of a massless fermion moving with the 
speed
\begin{equation}
v = \left\vert\frac{1 - \eta_0^2}{1 + \eta_0^2}\right\vert c \leq c,
\end{equation}
and coupled to a chemical potential
\begin{equation}
\mu = \frac{2 \eta_0}{1 + \eta_0^2} m c^2.
\end{equation}
The normalizability of the domain wall state requires $\varkappa > 0$
which again implies restrictions on $\vec p$. Only for $\eta_0 = 0$ or 
$\pm \infty$ there is no restriction and one obtains a relativistic massless 
left-handed domain wall fermion with $E = |\vec p \,| c$.

\section{Conclusions}

While the results presented here are easy to derive, they seem not to
constitute common knowledge in quantum mechanics. The theory of self-adjoint
extensions is not only mathematically elegant, but also physically relevant.
Therefore, we hope that our paper contributes to changing the view on
elementary textbook problems such as the particle in a box. While it may not
be appropriate to discuss the most general reflecting boundary condition in a
first encounter with the Schr\"odinger equation, one might at least point out
that other boundary conditions are possible as well. 

Since the main purpose of 
this paper is of conceptual nature, we have not made an attempt to determine 
the value of the self-adjoint extension parameter $\gamma$ for actual quantum 
dots. In typical cases, $\gamma$ may be very large, so that the wave function 
practically vanishes at the boundary. However, there may be specific situations
that lead to much smaller values of $\gamma$ and thus to quantitatively or
even qualitatively different behavior, such as bound states localized at the 
wall of a quantum dot. It would certainly be interesting to investigate this in 
more detail. As a special case of the general uncertainty relation for 
non-Hermitean operators, we have derived a generalized uncertainty relation for 
the self-adjoint position and the non-Hermitean momentum operator in a quantum 
dot. Interestingly, additional boundary terms enter this relation. In 
particular, negative energy states, which may seem to be inconsistent with the
uncertainty relation, are in perfect agreement with the generalized relation.
Minimal uncertainty wave packets, which saturate the corresponding inequality
and satisfy it as an equality, have been constructed as well. They are standard
Gaussian wave packets, but require very special boundary conditions and thus 
generically do not exist. If the self-adjoint extension parameter 
$\gamma(\vec x)$ vanishes everywhere at the boundary, a constant wave function
plays the role of a minimal uncertainty wave packet. Furthermore, we have shown 
that the spectrum depends monotonically on the self-adjoint extension parameter 
$\gamma$.

We have also applied the theory of self-adjoint extensions to theories of
relativistic fermions described by the Dirac equation. In $(1+1)$ and $(3+1)$
dimensions, we have considered the most general perfectly reflecting boundary
condition as a generalization of the one used in the MIT bag model. All these
boundary conditions necessarily explicitly break chiral symmetry. This can be 
avoided in the chiral bag model were the axial current is carried by a pion 
field outside the bag. We have also discussed generalized domain wall fermion
boundary conditions both in $(2+1)$ and in $(4+1)$ dimensions. The most general
perfectly reflecting boundary condition explicitly breaks $(1+1)$-d or $(3+1)$-d
Lorentz invariance or even 3-d spatial rotation invariance. A relativistic 
massless chiral fermion arises only for a particular choice of the self-adjoint 
extension parameters. For other values of the self-adjoint extension parameters,
one obtains a fermion at non-zero chemical potential, with a linear 
energy-momentum dispersion relation, however, with a speed $v < c$.

We conclude this paper by expressing our hope that it may contribute to 
emphasizing the sometimes subtle differences between Hermiticity and 
self-adjointness also in the teaching of quantum mechanics.

\section*{Acknowledgments}

We are indebted to F.\ Niedermayer and C.\ Tretter for very interesting
discussions. We also like to thank V.\ Maz'ya and B.\ Plamenevsky for hints to 
the mathematical literature. This work is supported in parts by the 
Schweizerischer Nationalfonds (SNF).

\end{document}